\documentclass{elsart}
\usepackage[pdftex]{graphicx}
\usepackage{amsmath}
\usepackage{hyperref}
\usepackage{color}

\newcommand\de[1]{\,{\mathrm d}#1} 
\newcommand\vek[1]{\mathbf{#1}}

\newcommand{\bmath}[1]{\mbox{\boldmath$#1$}}
\newcommand{\tenss}[1]{\bmath{#1}}                   % A second-order tensor
\newcommand{\tensf}[1]{\bmath{\mathsf{#1}}}          % A fourth-order tensor
\newcommand{\avge}[1]{\overline{#1}}
\newcommand{\trn}{{\sf ^T}}
\newcommand{\evek}[1]{\{\mathsf{#1}\}}
\newcommand{\emtrx}[1]{\left[\mathsf{#1}\right]}
\newcommand{\dcontr}{\,\colon}

\newcommand{\strain}{\varepsilon}
\newcommand{\stress}{\sigma}

\newcommand{\estress}{\lambda}

\newcommand{\scal}[1]{\mathnormal{#1}}

\newcommand{\uc}{\ensuremath \Omega } %Symbol for unit cell
\newcommand{\avg}[1]{\ensuremath \left\langle #1 \right\rangle} %Volume average 

\journal{International Journal for Multiscale Computational Engineering}

\begin{document}

\begin{frontmatter}

\title{Macroscopic constitutive law for Mastic Asphalt Mixtures from multiscale modeling}

\author[cideas]{Richard Valenta},
\ead{richard.valenta@centrum.cz}
\author[mech,cideas]{Michal \v{S}ejnoha\corauthref{auth}},
\ead{sejnom@fsv.cvut.cz}
\corauth[auth]{Corresponding author. Tel.:~+420-2-2435-4494;
fax~+420-2-2431-0775}
\author[mech]{Jan Zeman}
\ead{zemanj@cml.fsv.cvut.cz} 
\address[mech]{Department of Mechanics, Faculty of Civil
  Engineering, Czech Technical University in Prague, Th\' akurova 7,
  166 29 Prague 6, Czech Republic} 
\address[cideas]{Centre for Integrated Design of Advances Structures,
  Th\' akurova 7, 166 29 Prague 6, Czech Republic} 

\begin{abstract}
A well established framework of an uncoupled hierarchical modeling
approach is adopted here for the prediction of macroscopic material
parameters of the Generalized Leonov (GL) constitutive model intended
for the analysis of flexible pavements at both moderate and elevated
temperature regimes. To that end, a recently introduced concept of a
statistically equivalent periodic unit cell (SEPUC) is addressed to
reflect a real microstructure of Mastic Asphalt mixtures (MAm). While
mastic properties are derived from an extensive experimental program,
the macroscopic properties of MAm are fitted to virtual numerical
experiments performed on the basis of first order homogenization
scheme. To enhance feasibility of the solution of the underlying
nonlinear problem a two-step homogenization procedure is
proposed. Here, the effective material properties are first found for
a mortar phase, a composite consisting of a mastic matrix and a
fraction of small aggregates. These properties are then introduced in
place of the matrix in actual unit cells to give estimates of the
model parameters on macroscale. Comparison with the Mori-Tanaka
predictions is also provided suggesting limitations of classical
micromechanical models.
\end{abstract}


\end{frontmatter}

\section{Introduction}\label{sec:intro}
\begin{figure}[ht]
\begin{center}
\begin{tabular}{c@{\hspace{10mm}}c@{\hspace{10mm}}c}
\includegraphics[angle=0,width=3.45cm]{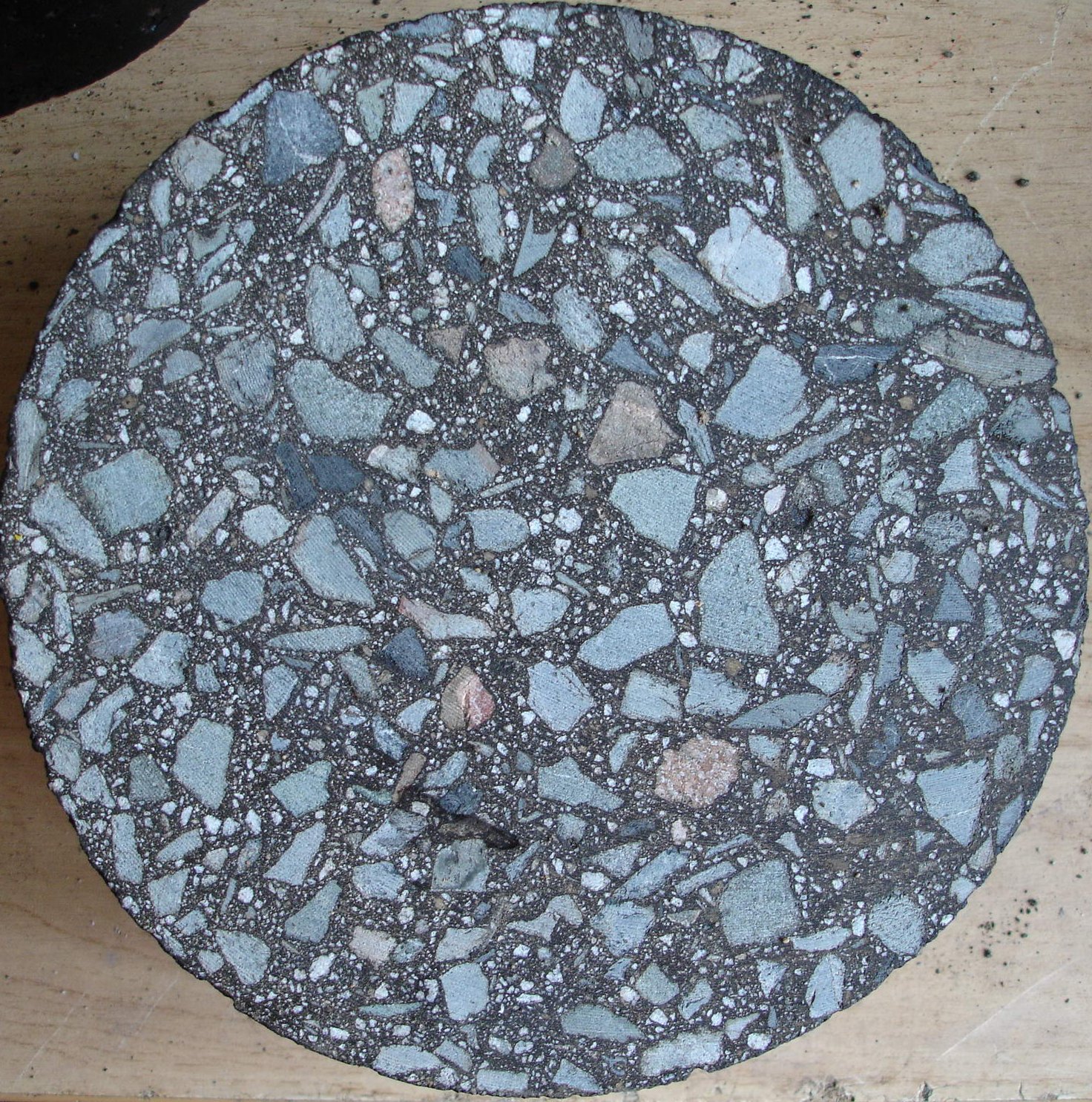}&
\includegraphics[angle=0,width=3.5cm]{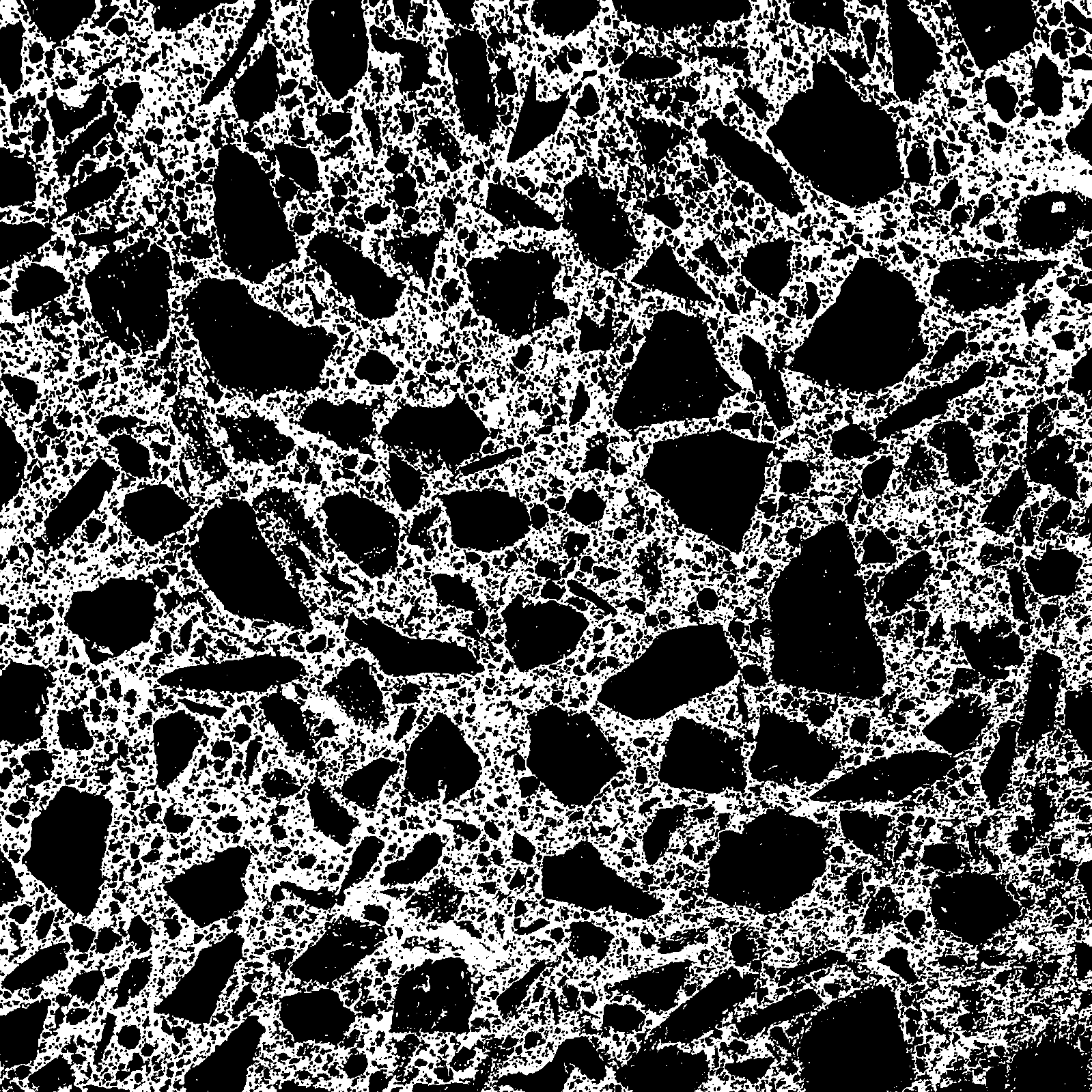}& 
\includegraphics[angle=0,width=3.5cm]{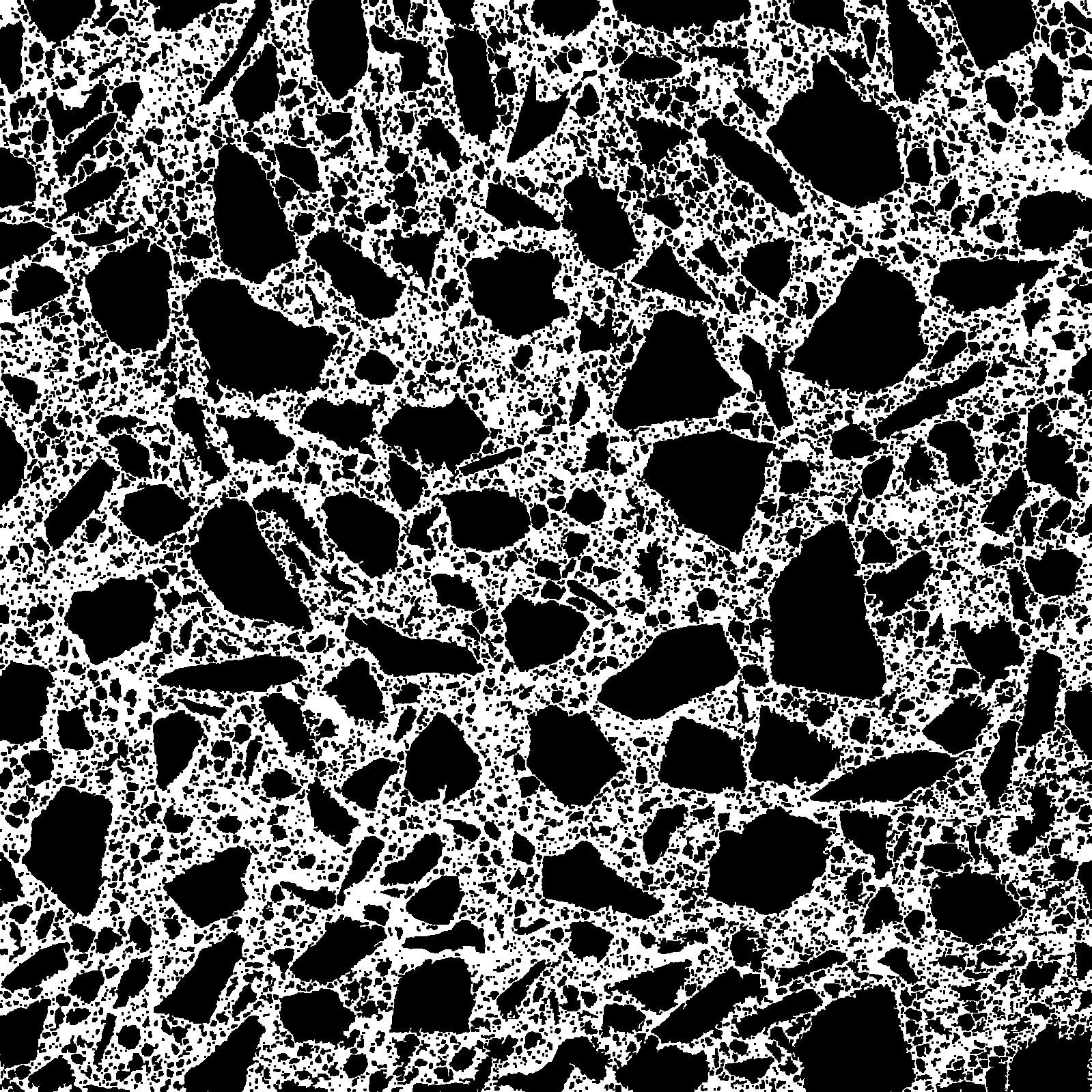}\\
(a)&(b)&(c)
\end{tabular}
\end{center}
\caption{(a) A real microstructure of an asphalt mixture, (b) Original binary image, (c) Improved binary image}
\label{F:intro:1}
\end{figure}

%%%%%%%%%%%%%%%%%%%%%%%%%%%%%%%%%%%%%%%%%%%%%%%%%%%%%%%%%%%%%%%%%%%%%%%%%%%%%
As seen in Fig.~\ref{F:intro:1}(a), asphalt mixtures represent in general a
highly heterogeneous material with complex microstructure consisting
at minimum of mastic binder, aggregates and voids. When limiting our
attention to Mastic Asphalt mixtures, used typically in traffic
arteries of substantial importance, the fraction of voids becomes
negligible.  A binary image of such a two-phase material system
plotted in Fig.~\ref{F:intro:1}(b) is then readily available. The literature
offers several distinct routes taking advantage of such a
representation. 

%%%%%%%%%%%%%%%%%%%%%%%%%%%%%%%%%%%%%%%%%%%%%%%%%%%%%%%%%%%%%%%%%%%%%%%%%%%%%%%%%%%%%%%%%%%%%%
%%%%%%%%%%%%%%%%%%%%%%%%%%%%%%%%%%%%%%%%%%%%%%%%%%%%%%%%%%%%%%%%%%%%%%%%%%%%%%%%%%%%%%%%%%%%%%

Phenomenological constitutive models enhanced by introducing a
microstructure linked aggregate distribution
tensor~\cite{Masad:MM:2005,Panoskaltsis:2005} were proposed, following
the footsteps of isotropic viscoelastic or viscoplastic models
developed since the early 1990s~\cite[to cite a
  few]{Sousa:TRR:1993,Sousa:JAAPT:1994,Weissman:2003,Huang:JMCE:2004,Fang:FEAD:2004},
to account for an intrinsic aggregate structure anisotropy.  Recently,
detailed micromechanical models developed on the basis of image
processing of either two-dimensional digital photography or
three-dimensional X-ray computed tomography have been favored for the
performance assessment of asphalt concrete composites. Typically, a
sufficiently large sample of microstructure of a two-phase composite
system, transformed into a binary image, was adopted to serve as a
representative volume element (RVE). Either direct application of
these models~\cite{Papagiannakis:TRR:2002,Abbas:JMCE:2004,You:MS:2009}
or application of simplified artificial microstructures of the same
type~\cite{Dai:IJNAMG:2006} was examined to provide estimates of the
macroscopic response. While capturing both the local and overall
behavior sufficiently accurately, most of these models suffer from
computational inefficiently. Exploitation of the true potential of
classical averaging micromechanical schemes combined with
``bottom-up'' hierarchical homogenization technique thus appears
rather natural.  As an example we recall a successful application of
the Mori-Tanaka method in upscaling viscoelastic properties of asphalt
mixtures at low
temperatures~\cite{Lackner:PECT:2004,Lackner:JMCE:2005,Lackner:CM:2006,Lackner:ACPSEM:2007}.

Growing from the roots planted by the authors in the early
2000s~\cite{Matous:2000:GEI,Zeman:2001:EPG,Sejnoha:2002:OVR} the
present contribution combines these various aspects of micromechanical
modeling into a relatively simple, yet reliable and efficient
computational scheme. While still incorporating the prominent
``bottom-up'' uncoupled multiscale homogenization scheme, the
effective properties on individual scales are found as volume averages
of local stress and strain fields developed inside the so-called
Statistically Equivalent Periodic Unit Cell~\cite{Zeman:MSMSE:2007}.

Rendering the desired macroscopic constitutive model that describes
the homogenized response of MAm to general loading actions thus
endeavors to the formulation of a suitable micromechanical model on
individual scales and to associated experimental work being jointly
the building blocks of the upscaling procedure. Three particular
scales shown in Fig.~\ref{F:intro:2} are considered in the present
study:

%%%%%%%%%%%%%%%%%%%%%%%%%%%%%%%%%%%%%%%%%%%%%%%%%%%%%%%%%%%%%%%%%%%%%%%%%%%%%%%%%%%%%%%%%%%%%%
\begin{figure}[ht]
\begin{center}
\includegraphics[angle=0,width=9cm]{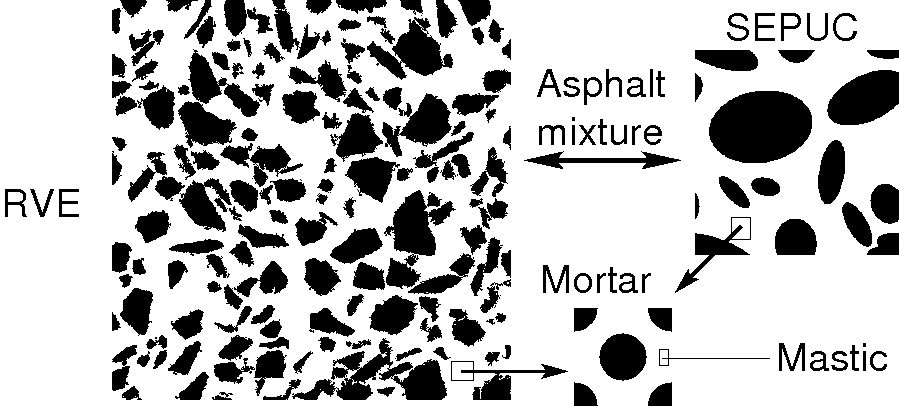}
\end{center}
\caption{Three distinct scales of Mastic Asphalt mixture}
\label{F:intro:2}
\end{figure}
%%%%%%%%%%%%%%%%%%%%%%%%%%%%%%%%%%%%%%%%%%%%%%%%%%%%%%%%%%%%%%%%%%%%%%%%%%%%%%%%%%%%%%%%%%%%%%

\begin{itemize}
\item Although the mastic-phase itself is a composite consisting of a
  filler and a bituminous binder, it is assumed in the present study
  to be well represented by a temperature and rate dependent
  homogeneous isotropic material. Since limiting our attention to
  elevated temperatures exceeding 30$^0$C, the GL model discussed in
  Section~\ref{bsec:leonov} is exploited to provide for experimentally
  observed nonlinear viscoelastic behavior of bituminous matrices. The
  required experimental program to calibrate the model parameters is
  outlined in Section~\ref{bsec:leonov-mastic}.

\item As presented in Section~\ref{bsec:BRVE}, the mortar-phase
  naturally arises through the process of removing aggregates from
  original microstructure, Fig.~\ref{F:intro:1}(b), passing 2.26 mm
  sieve. Introducing the principal assumption of the adopted upscaling
  procedure - the GL model is suitable for governing the homogenized
  response on every scale - allows us to derive the model parameters
  on the mortar-scale by running a set of virtual numerical
  experiments on the simplified periodic hexagonal array model plotted
  in Fig.~\ref{F:intro:2}. The selection of this geometrical representation
  of the mortar composite is purely an assumption building upon the
  conclusion that at least for the mastic-phase and low-temperature
  creep the response is independent of the filler shape and
  mineralogy~\cite{Lackner:JMCE:2005}. Various computational
  strategies delivering the searched macroscopic response are
  presented in Sections~\ref{bbsec:MT}~-~\ref{bbsec:FEM}.
\item The SEPUC, also seen in Fig.~\ref{F:intro:2}, is selected to
  predict the macroscopic response of MAm, where large aggregates are
  bonded to a mortar-phase being homogeneous and isotropic. The
  elliptical shape of aggregates has already been successfully used in
  detailed micromechanical simulations presented
  in~\cite{Dai:IJNAMG:2006}.  The issue of constructing SEPUC for the
  class of Mastic Asphalt mixtures is briefly addressed in
  Section~\ref{bsec:sepuc}.
\item The proposed upscaling procedure consistent with
  Fig.~\ref{F:intro:2} is then applied in
  Sections~\ref{bsec:leonov-mortar} and~\ref{bsec:leonov-MAM} to
  deliver the necessary parameters of the homogenized GL
  model. 
%Whether this approach can be brought to points of application
%is partially addressed in Section~\ref{sec:MSA}.
\end{itemize}

In the following text, $\vek{a}$, $\tenss{a}$ and $\tensf{A}$ denote a
vector, a symmetric second-order and a fourth-order tensor,
respectively. The standard summation notation is adopted, i.e.,
$\tensf{A}\dcontr\tensf{B}$ and $\tensf{A}\dcontr\tenss{b}$ denote the
sums $A_{ijkl}B_{klrs}$ and $A_{ijkl}b_{kl}$,
$\tenss{a}\dcontr\tenss{b}$ being $a_{ij}b_{ij}$ represents a scalar
quantity and $\tenss{a}\cdot\vek{b}$ stands for $a_{ij}b_j$, where
the summation with respect to repeated indices is used. The symbol
$\evek{a}$ is reserved for a column matrix or a vectorial
representation of symmetric second-order tensor while the notation
$\emtrx{L}$ is employed for a matrix representation of a fourth-order
tensor~\cite{Bittnar:1996:NMM}.

\section{Concept of statistically equivalent periodic unit cell}\label{sec:SEPUC}
%%%%%%%%%%%%%%%%%%%%%%%%%%%%%%%%%%%%%%%%%%%%%%%%%%%%%%%%%%%%%%%%%%%%%%%%%%%%%%%%%%%%%%%%%%%%%%%%%%%%%%%%%%%%%%%%%%%%%%%%%%%%%

It has been demonstrated in our previous work, see
e.g.~\cite{Zeman:2001:EPG,Sejnoha:MCE:2004,Sejnoha:SEM:2008} that
image analysis of real, rather then artificial, material systems plays
an essential role in the derivation of a reliable and accurate
computational model. This issue is revisited here to get, although
considerably simplified, a reasonably accurate computational model
(RVE$\rightarrow$SEPUC) representing the real microstructure of Mastic
Asphalt mixtures. 

\subsection{From real microstructure to computationally feasible RVE}\label{bsec:BRVE}
%%%%%%%%%%%%%%%%%%%%%%%%%%%%%%%%%%%%%%%%%%%%%%%%%%%%%%%%%%%%%%%%%%%%%%
Notice a 1,600$\times$1,600 pixel resolution binary image in
Fig.~\ref{F:intro:1}(b) created through the process of color
segmentation of the original image in Fig.~\ref{F:intro:1}(b) using the
freeware graphical software GIMP. Visible flaws, e.g. defects inside
large aggregates, required, however, further modification. To that
end, an automated numerical tool was developed to determine a
boundary of each aggregate, fill defects inside large aggregates and
separate all aggregates as much as possible. The resulting improved
image seen in Fig.~\ref{F:intro:1}(c) then allows for microstructure
quantification in terms of various statistical descriptors.

%%%%%%%%%%%%%%%%%%%%%%%%%%%%%%%%%%%%%%%%%%%%%%%%%%%%%%%%%%%%%%%%%%%%%%%%%%%%%%%%%%%%%%%%%%%%%%
\begin{figure}[ht]
\begin{center}
\includegraphics[angle=0,width=7cm]{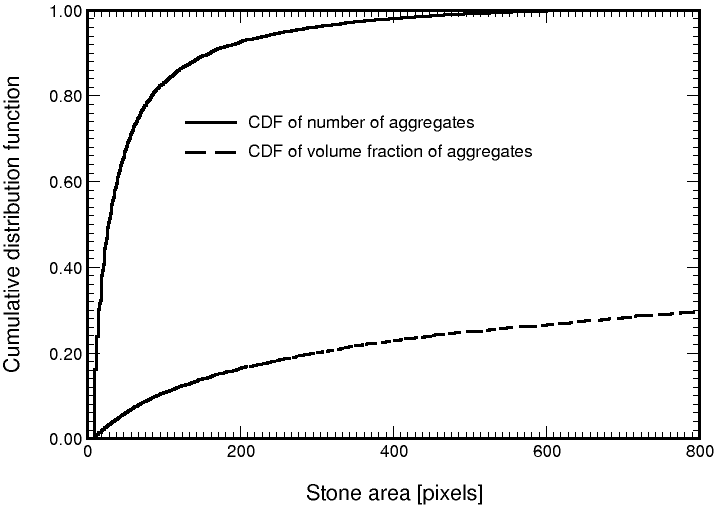}
\end{center}
\caption{Cumulative distribution function}
\label{F:rve:1}
\end{figure}
%%%%%%%%%%%%%%%%%%%%%%%%%%%%%%%%%%%%%%%%%%%%%%%%%%%%%%%%%%%%%%%%%%%%%%%%%%%%%%%%%%%%%%%%%%%%%%

\begin{figure}[ht]
\begin{center}
\begin{tabular}{c@{\hspace{10mm}}c}
\includegraphics[angle=0,width=4.cm]{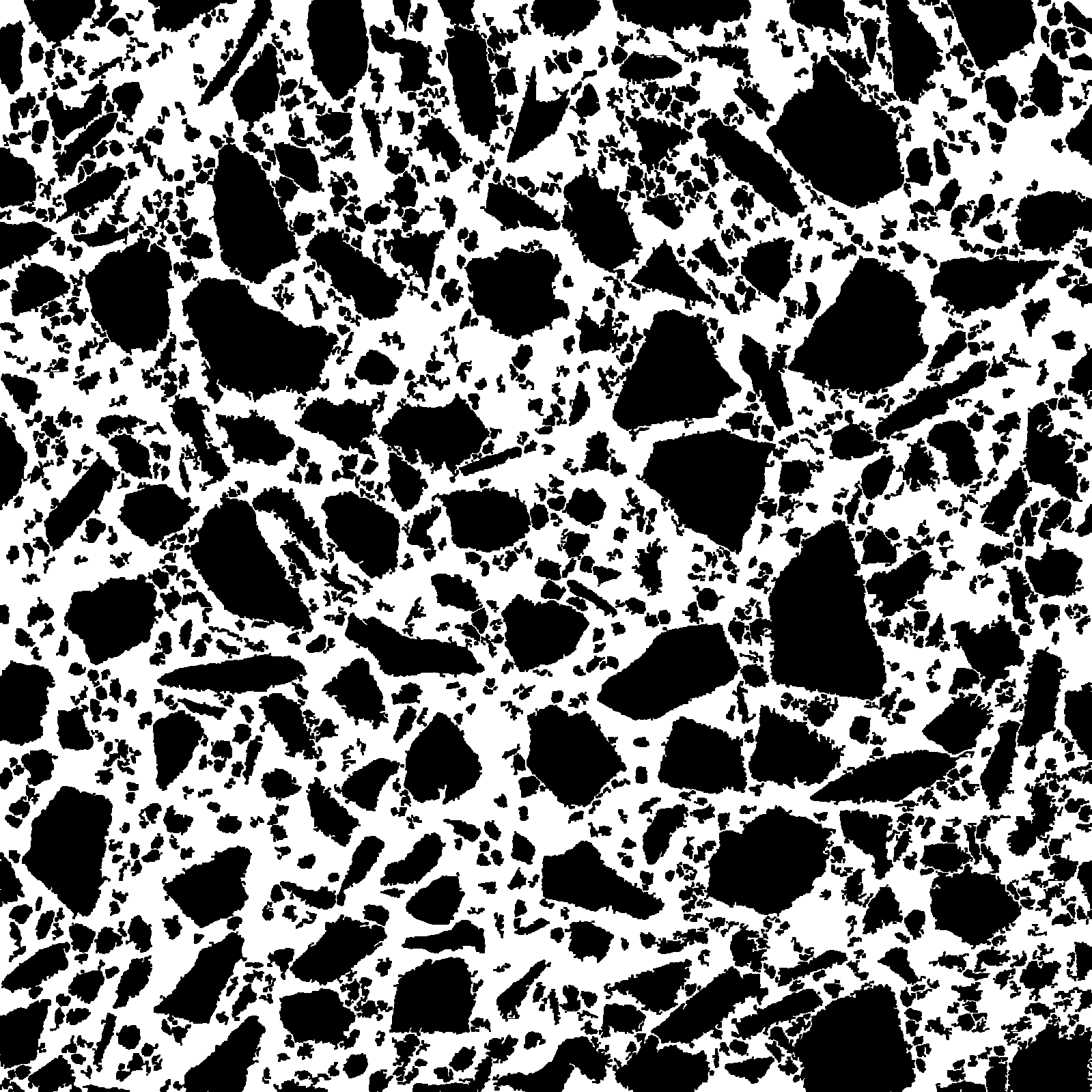}&
\includegraphics[angle=0,width=4.cm]{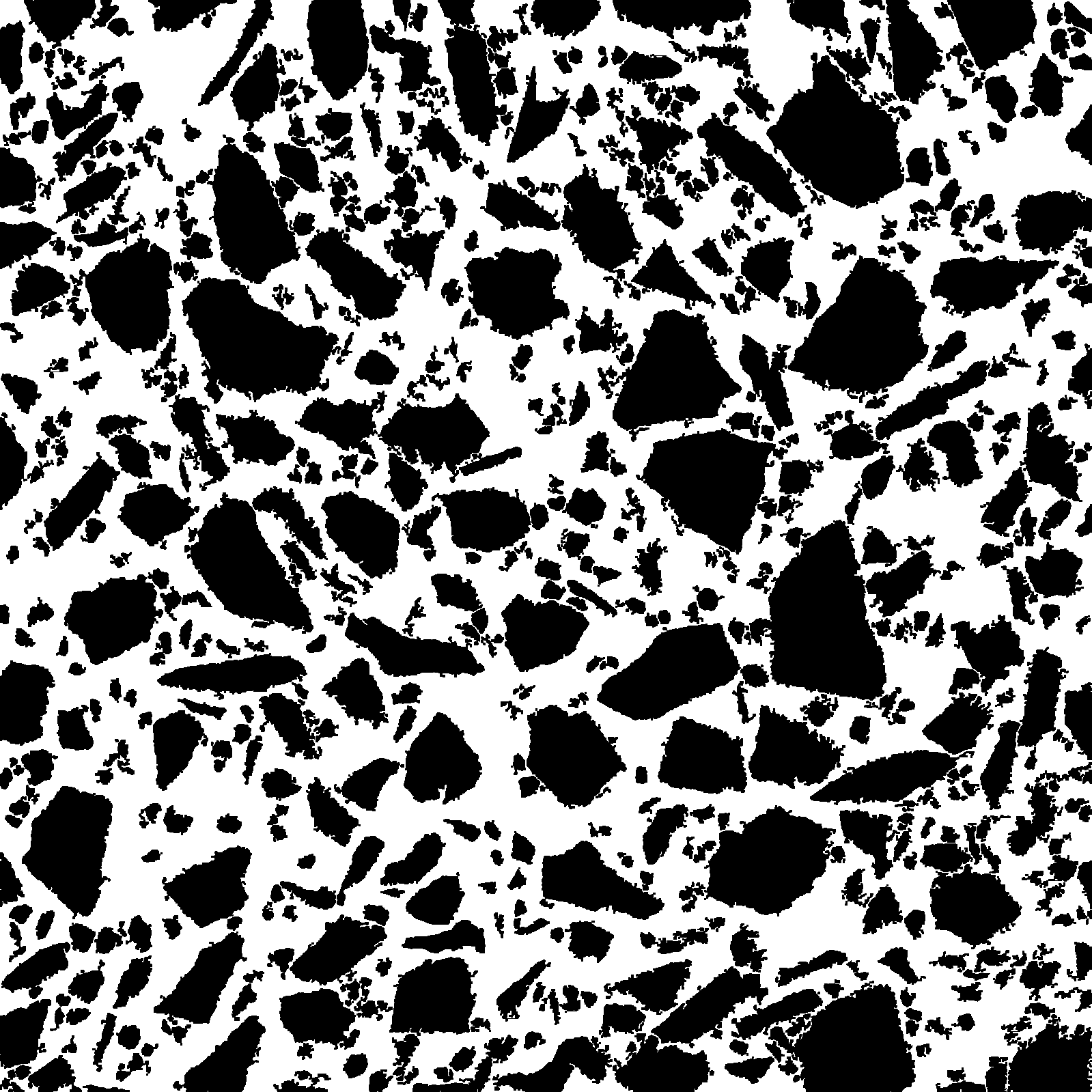}\\
(a)&(b)\\
\includegraphics[angle=0,width=4.cm]{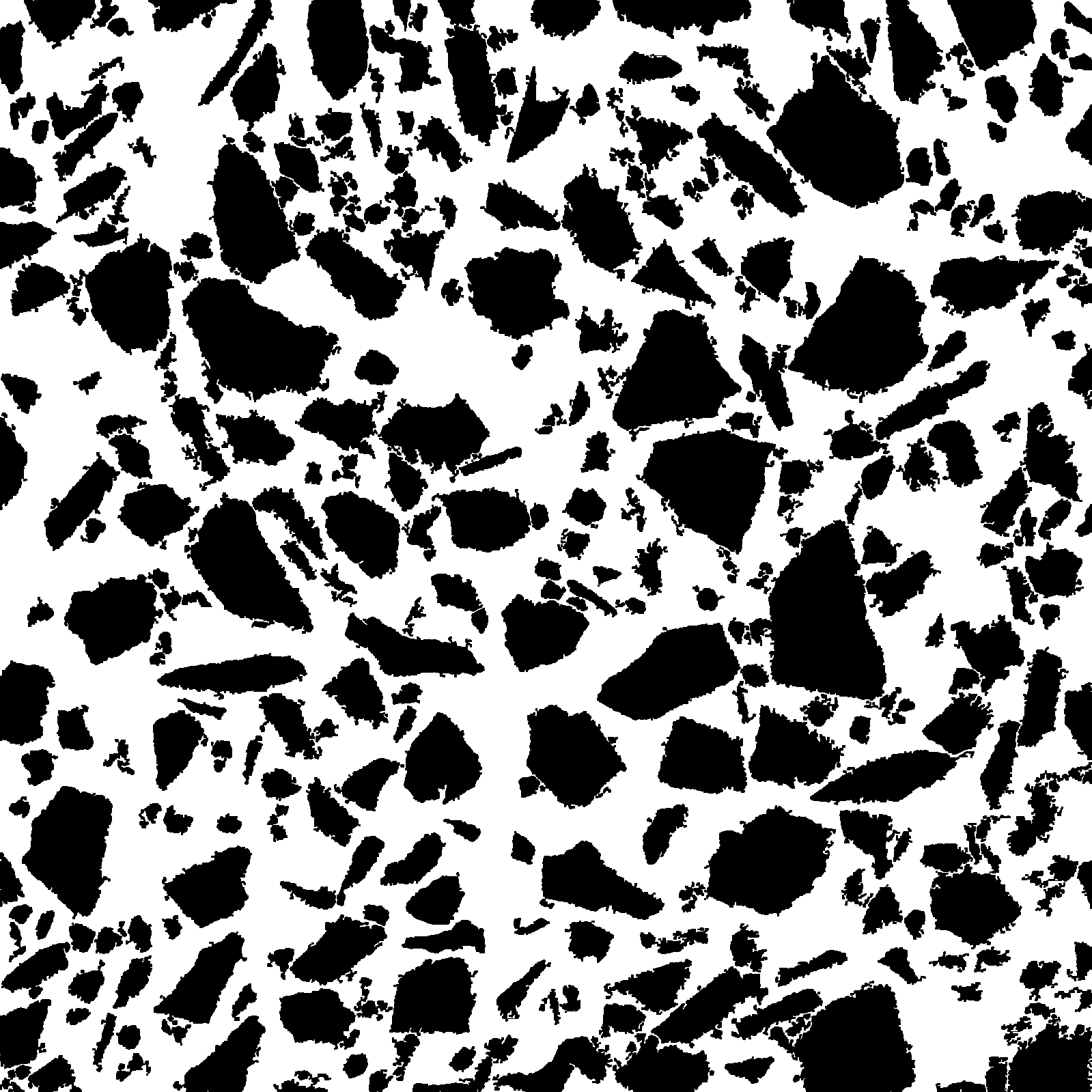}&
\includegraphics[angle=0,width=4.cm]{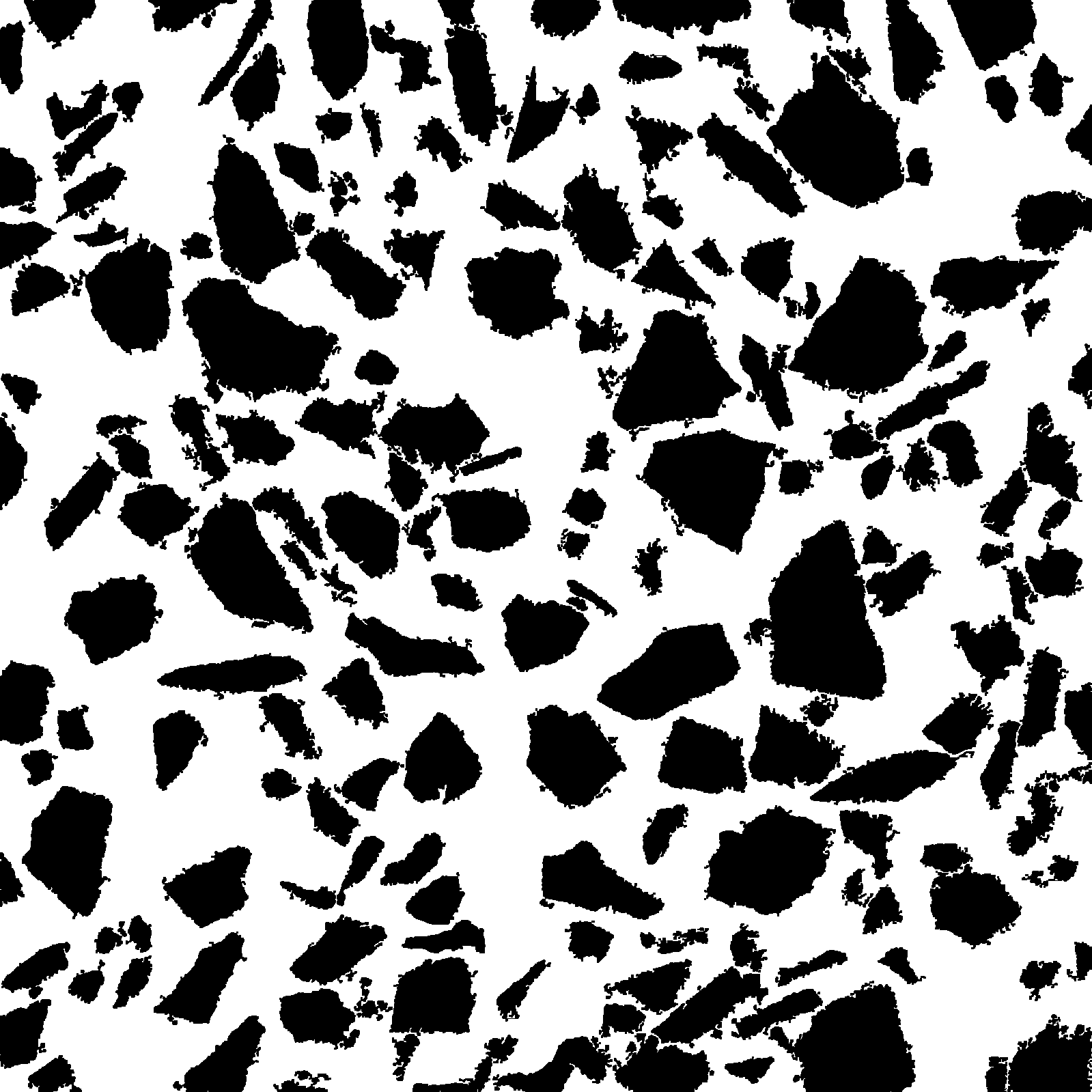}\\
(c)&(d)
\end{tabular}
\end{center}
\caption{Examples of binary images of original microstructure after eliminating stone fragments smaller than (in area):
(a) 150px, (b) 300px, (c) 600px, (d) 1200px}
\label{F:rve:2}
\end{figure}

To open this subject, we consider first cumulative distribution
functions of aggregates plotted in Fig.~\ref{F:rve:1}. The two graphs
suggest a relatively large amount of small stones (cumulative
distribution function of number of aggregates) which might seem, at
least from their volume fraction point of view (cumulative
distribution function of volume fraction of aggregates), almost
negligible. Eliminating the small aggregates to simplify the original
microstructure is naturally promoted. Several examples of such
microstructures appear in Fig.~\ref{F:rve:2}. If admitting removal of
all fragments smaller than 1,200 pixels in area (aggregates passing
2.36mm sieve) we reduce the total number of stones by 97\% while the
original volume fraction of aggregates dropped down only by 21\%.
However, neglecting the small aggregates completely would severely
underestimate the final macroscopic predictions. A two-scale
homogenization approach is therefore inevitable.

%%%%%%%%%%%%%%%%%%%%%%%%%%%%%%%%%%%%%%%%%%%%%%%%%%%%%%%%%%%%%%%%%%%%%%%%%%%%%%%%%%%%%%%%%%%%%%
%%%%%%%%%%%%%%%%%%%%%%%%%%%%%%%%%%%%%%%%%%%%%%%%%%%%%%%%%%%%%%%%%%%%%%%%%%%%%%%%%%%%%%%%%%%%%%

\subsection{Upscaling material properties towards Mastic Asphalt Mixes}\label{bsec:MAM}
%%%%%%%%%%%%%%%%%%%%%%%%%%%%%%%%%%%%%%%%%%%%%%%%%%%%%%%%%%%%%%%%%%%%%%

\begin{figure}[ht]
\begin{center}
\begin{tabular}{c@{\hspace{5mm}}c@{\hspace{5mm}}c@{\hspace{5mm}}c}
\includegraphics[angle=0,width=2.cm]{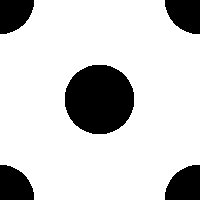}&
\includegraphics[angle=0,width=2.cm]{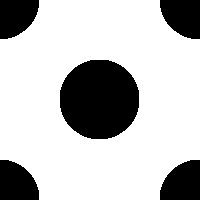}&
\includegraphics[angle=0,width=2.cm]{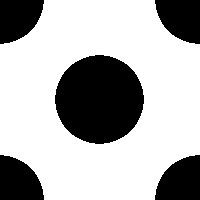}&
\includegraphics[angle=0,width=2.cm]{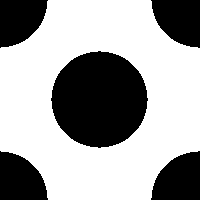}\\
(a) $A_{s} = 0.19A$ &(b) $A_{s} = 0.25A$ &(c) $A_{s} = 0.31A$ &(d) $A_{s} = 0.34A$
\end{tabular}
\end{center}
\caption{PUC of mortar phase containing stone fragments smaller than (in area):
(a) 150px, (b) 300px, (c) 600px, (d) 1200px}
\label{F:rve:3}
\end{figure}

Suppose that the microstructure in Fig.~\ref{F:rve:2}(d) is sufficient
to provide reasonably accurate predictions of macroscopic response of
MAm. With reference to Fig.~\ref{F:intro:2} and to findings presented
in~\cite{Lackner:JMCE:2005} we further assume the mortar phase to be
well represented by a periodic hexagonal arrangement of aggregates of
a circular cross-section. The resulting Mastic Asphalt mixture then
consists of 41\% volume of large aggregates and 59\% volume of mortar
phase, whereas the mortar-mastic composite was calculated to have 34\%
volume of small stones and 64\% volume of mastic phase. The latter
composite is shown in Fig.~\ref{F:rve:3}(d). For further reference the
subscript $s$ will be reserved for aggregates (stones) whereas the
subscript $m$ will be used to denote the binder (either mortar or
mastic matrix).

%%%%%%%%%%%%%%%%%%%%%%%%%%%%%%%%%%%%%%%%%%%%%%%%%%%%%%%%%%%%%%%%%%%%%%%%%%%%%%%%%%%%%%%%%%%%%%
%%%%%%%%%%%%%%%%%%%%%%%%%%%%%%%%%%%%%%%%%%%%%%%%%%%%%%%%%%%%%%%%%%%%%%%%%%%%%%%%%%%%%%%%%%%%%%

Given the geometrical models for individual scales now opens the way
to the derivation of the effective elastic properties of MAm employing
the uncoupled ``bottom-up'' upscaling scheme. While confirming
applicability of this approach, these results will further serve as a
point of departure for generating the statistically equivalent
periodic unit cells and at the same time will provide a source of data
to check the quality of the results obtained for these artificial
microstructures, see Section~\ref{bsec:sepuc}. 

A number of homogenization techniques is available in the literature.
Three particular representatives will be now briefly reviewed. Owing
to the solution of the underlying nonlinear viscoelastic problem, the
formulation will be presented in an incremental format.

\subsubsection{The Mori-Tanaka method}\label{bbsec:MT}  
%%%%%%%%%%%%%%%%%%%%%%%%%%%%%%%%%%%%%%%%%%%%%%%%%%%%%%%%%%%%%%%%%%%%%%
Consider a two-phase composite medium $\Omega$ loaded on its outer
boundary $S$ by an affine displacement field
$\Delta\tenss{u}_0=\Delta\tenss{E}\cdot\vek{x}$ consistent with the
macroscopic uniform strain increment $\Delta\tenss{E}$. The local
constitutive equation can be written as
\begin{equation}
\Delta\tenss{\stress}(\vek{x}) = \tensf{L}(\vek{x})\dcontr\Delta\tenss{\strain}(\vek{x})+\Delta\tensf{\lambda}(\vek{x})
\hspace{0.5cm}\text{or}\hspace{0.5cm}
\Delta\tenss{\stress}_r = \tensf{L}_r\dcontr\Delta\tenss{\strain}_r+\Delta\tensf{\lambda}_r,\label{eq:MT-1}
\end{equation}
when limiting our attention to a piecewise uniform variation of the
stress and strain fields within individual phases; $\tensf{L}_r$ is
then the stiffness tensor of the phase $r=s,m$ and the increment of
eigenstress $\Delta\tensf{\lambda}_r$ represents contributions caused
by various non-mechanical physical sources. In the present study we
admit only the creep strains developed in the binder phase thus
leaving the aggregates to remain linearly elastic. In the framework of
Dvorak's Transformation Field
Analysis~\cite{Dvorak92a,Dvorak:1992:TFA} the local strain increments
then read
\begin{eqnarray}
\Delta\tenss{\strain}_s &=& \tensf{A}_s\dcontr\Delta\tenss{E}+\tensf{D}_{sm}\dcontr\Delta\tenss{\mu}_m,\\
\Delta\tenss{\strain}_m &=& \tensf{A}_m\dcontr\Delta\tenss{E}+\tensf{D}_{mm}\dcontr\Delta\tenss{\mu}_m,
\end{eqnarray}
where $\tensf{A}_r,\tensf{D}_{rm}, r=s,m$, are the mechanical strain
localization tensors and transformation influence functions,
respectively. The phase eigenstrain $\tenss{\mu}_m$ is introduced to
account for a nonlinear viscoelastic behavior of the binder. Note that
for a two-phase medium, the transformation influence functions are
directly available in terms of the strain localization
tensors~\cite{Dvorak92a}.

Next, writing the macroscopic constitutive law as
\begin{equation}
\Delta\tenss{\Sigma} = \tensf{L}^{\text{hom}}\dcontr\Delta\tenss{E}+\Delta\tenss{\Lambda},
\end{equation}
and realizing that
\begin{equation}
\Delta\tenss{\Sigma}=c_s\Delta\tenss{\sigma}_s+c_m\Delta\tenss{\sigma}_m,
\end{equation}
yields the macroscopic stiffness matrix and the macroscopic
eigenstress~\cite{Dvorak92a} as
\begin{eqnarray}
\tensf{L}^{\text{hom}}  &=& c_s\tensf{L}_s\dcontr\tensf{A}_s+c_m\tensf{L}_m\dcontr\tensf{A}_m,\\
%\Delta\tenss{\Lambda} &=& c_s\left(\tensf{L}_s\dcontr\tensf{D}_{sm}\right)\dcontr\Delta\tenss{\mu}_m+
%c_m\left[\tensf{L}_m\dcontr\left(\tensf{D}_{mm}-\tensf{I}\right)\right]\dcontr\Delta\tenss{\mu}_m.
\Delta\tenss{\Lambda} &=& -\underbrace{\left[c_s\left(\tensf{L}_s\dcontr\tensf{D}_{sm}\dcontr\tensf{M}_m\right)+
  c_m\left(\tensf{L}_m\dcontr\tensf{D}_{mm}\dcontr\tensf{M}_m\right)\right]}_{=0}\dcontr\Delta\tenss{\lambda}_m+c_m\Delta\tenss{\lambda}_m,\nonumber\\
 &=& c_m\tenss{\lambda}_m \,=\,-c_m\tensf{L}_m\tenss{\mu}_m.
\end{eqnarray}
Herein, the strain localization tensors $\tensf{A}_r$ are found from
the Mori-Tanaka (MT) method~\cite{Mori:1973:MTM}. In Benveniste's
reformulation~\cite{Benvensite:1987:MTM}, the method approximates the
particle interactions by loading an isolated heterogeneity bonded to
an unbounded matrix by a uniform stress or strain found in the
matrix. Introducing the partial strain concentration (localization)
tensor $\tensf{T}_s$ for the phase $s$ (note that $\tensf{T}_m
=\tensf{I}$, the identity tensor), the local strain increment in
aggregates is, in the absence of inelastic effects, provided by
\begin{equation}
\Delta\tenss{\strain}_s = \tensf{T}_s\dcontr\Delta\tenss{\strain}_m.
\end{equation}
Next, writing $\Delta\tenss{E}=c_s\Delta\tenss{\strain}_s+c_m\Delta\tenss{\strain}_m$ readily provides
\begin{eqnarray}
\Delta\tenss{\strain}_m&=&\left[c_m\tensf{I}+c_s\tensf{T}_s\right]^{-1}\dcontr\Delta\tenss{E},\\
\Delta\tenss{\strain}_s&=&\left(\tensf{T}_s\dcontr\tensf{A}_m\right)\dcontr\Delta\tenss{E},
\end{eqnarray}
which gives the strain localization tensors in the form
\begin{equation}
\tensf{A}_m=\left[c_m\tensf{I}+c_s\tensf{T}_s\right]^{-1}\hspace{1cm}\tensf{A}_s\,=\,\tensf{T}_s\dcontr\tensf{A}_m.
\end{equation}
Closed form solutions for the localization tensor $\tensf{T}_s$ are
provided in the literature for certain types of heterogeneities, see
e.g.~\cite{Mura:1982:MDS}. Here, the model of aligned infinite
cylinders of circular cross-sections, consistent with the assumed
plane strain conditions, is adopted. Since used later in the Appendix,
%Section~\ref{bbsec:MT-visco}, 
we present here the homogenized in-plane shear modulus $m$ in the form
\begin{equation}
m_{\sf{MT}}=\frac{m_sm_m(k_m+2m_m)+k_mm_m(c_sm_s+c_mm_m)}{k_mm_m+(k_m+2m_m)(c_sm_m+c_mm_s)},\label{eq:m-hom-MT}
\end{equation}
where $m_r, r=s,m$ are corresponding shear moduli of individual
phases. If both phases are also isotropic we get
$k_r=m_r/(1-2\nu_r)$, where $\nu_r$ is the Poisson ratio.
Finally note that all localization and transformation tensors
mentioned in this section are functions of the instantaneous material
properties of the binder phase, the current value of the viscoelastic
shear modulus in particular.

\subsubsection{The Fast Fourier Transform method}\label{bbsec:FFT}  
%%%%%%%%%%%%%%%%%%%%%%%%%%%%%%%%%%%%%%%%%%%%%%%%%%%%%%%%%%%%%%%%%%%%%%
The Fast Fourier Transform (FFT) method proposed
in~\cite{Moulinec:NCCM:99} can be imagined as a passage between
classical micromechanics models and the concept of periodic
homogenization. To introduce this approach consider again a two-phase
composite medium subjected to the same boundary conditions as in the
previous section. 

%%%%%%%%%%%%%%%%%%%%%%%%%%%%%%%%%%%%%%%%%
\begin{figure}[ht]
\begin{center}
\includegraphics[angle=0,width=12cm]{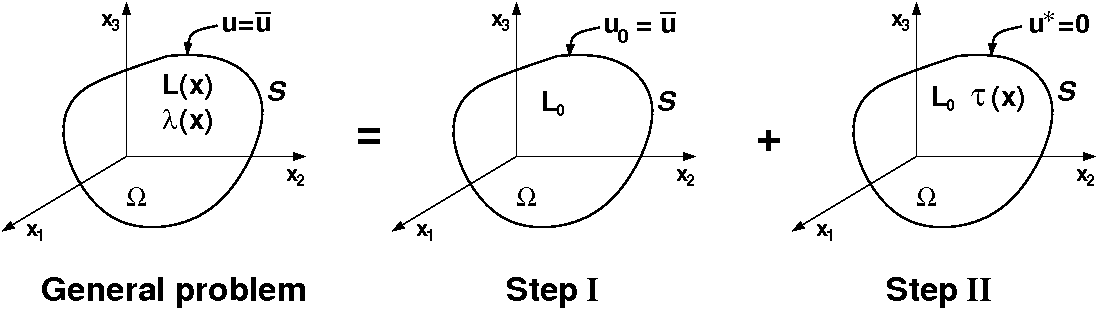}
\end{center}
\caption{Body with prescribed surface displacements including eigenstresses}
\label{F:rve:4}
\end{figure}
%%%%%%%%%%%%%%%%%%%%%%%%%%%%%%%%%%%%%%%%%

As suggested by Hashin and Shtrikman \cite{Hashin:1962a} the local
stress and strain fields in Eq.~\eqref{eq:MT-1}$_1$ can be found from
two auxiliary boundary value problems, Fig.~\ref{F:rve:4}. The
procedure starts by assuming a geometrically identical body with a
certain reference homogeneous, but generally anisotropic, medium
$\tensf{L}_0$ and the same prescribed displacements. The corresponding
uniform strain $\tenss{E}$ and stress $\tenss{\Sigma}$ fields are
related through constitutive law as
\begin{equation}\label{ec41}
\Delta\tenss{\Sigma}= \tensf{L}_0\dcontr\Delta\tenss{E} \,\,\text{in}\,\, \Omega,\,\,\Delta\tenss{u}_0 = \Delta\avge{\tenss{u}} \,\,\text{on}\,\,S. 
\end{equation}
Following the Hashin-Shtrikman idea, we introduce the symmetric
stress polarization tensor $\tenss{\tau}$ such that
\begin{equation}\label{ec43}
\Delta\tenss{\sigma}(\vek{x}) = \tensf{L}_0\dcontr\Delta\tenss{\strain}(\vek{x}) +\Delta\tenss{\tau}(\vek{x}).
\end{equation}
In addition, denote 
\begin{equation}\label{ec44}
\tenss{u}^* = \tenss{u} - \tenss{u}_0 \,\, {\rm in} \,\,\Omega  , \tenss{u}^* =
\vek{0} \,\, {\rm on} \,\, S,\,\,\tenss{\strain}^* = \tenss{\strain} - \tenss{E}, \,\,\text{in}\,\,\Omega.
\end{equation}
The increment of stress polarization tensor $\tenss{\tau}$ then becomes
\begin{equation}
\Delta\tenss{\tau}(\vek{x})=\left(\tensf{L}(\vek{x})-\tensf{L}_0\right)\dcontr\Delta\tenss{\strain}(\vek{x})-\Delta\tenss{\lambda}(\vek{x}).\label{eq:tau}
\end{equation}
If the polarization stress is known, the fluctuation part of the
strain field $\tenss{\strain}^*$ can be obtained via Green's function
$\tensf{\Gamma}^0$ for a given reference medium in the form, see
e.g.~\cite{Willis:1977:BSCE,Moulinec:NCCM:99},
\begin{equation}\label{eq:strain_green}
\Delta\tenss{\strain}^*(\vek{x})=-\int_\Omega\tensf{\Gamma}^0(\vek{x}-\vek{x}')\dcontr\Delta\tenss{\tau}(\vek{x}')\de{\vek{x}'}.
\end{equation}
Combining Eqs.~\eqref{ec44}$_3$,\eqref{eq:tau}
and~\eqref{eq:strain_green} we obtain the so called {\em periodic
  Lippmann Schwinger} integral equation for a given reference medium
as
\begin{equation}
\Delta\tenss{\strain}(\vek{x})+\int_\Omega\tensf{\Gamma}^0(\vek{x}-\vek{x}')\dcontr
\left[(\tensf{L}(\vek{x}')-\vek{L}_0)\dcontr\Delta\tenss{\strain}(\vek{x}')-\Delta\tenss{\lambda}(\vek{x})\right]\de{\vek{x}'}=\Delta\tenss{E}.
\end{equation}
This equation can be solved by the following iterative procedure
\begin{equation}
\Delta\tenss{\strain}^{k+1}(\vek{x})=\Delta\tenss{E}-\int_\Omega\tensf{\Gamma}^0(\vek{x}-\vek{x}')\dcontr
\left[(\tensf{L}(\vek{x}')-\tensf{L}^0)\dcontr\Delta\tenss{\strain}^{k}(\vek{x}')-\Delta\tenss{\lambda}^{k}(\vek{x}')\right]\de{\vek{x}'}.
\end{equation}
Typically, the Fast Fourier Transform (or rather its discrete version
when dealing directly with binary images) is employed to solve the
above equation. Details on actual numerical implementation can be
found in~\cite{Moulinec:NCCM:99}. In the absence of inelastic effects
the above equation simplifies as
\begin{equation}
\tenss{\strain}^{k+1}(\vek{x})=\tenss{E}-\int_\Omega\tensf{\Gamma}^0(\vek{x}-\vek{x}')\dcontr
\left[(\tensf{L}(\vek{x}')-\tensf{L}^0)\dcontr\tenss{\strain}^{k}(\vek{x}')\right]\de{\vek{x}'}.
\end{equation}
For plane-strain problem the homogenized elastic stiffness matrix
$\tensf{L}^{\text{hom}}$ can be found from the solution of four
successive elasticity problems. To that end, the RVE is loaded, in
turn, by each of the four components of $\tenss{E}$, while the other
three components vanish. The volume stress averages normalized with
respect to $\tenss{E}$ then furnish individual columns of
$\tensf{L}^{\text{hom}}$.

\subsubsection{First order homogenization of periodic fields}\label{bbsec:FEM}  
%%%%%%%%%%%%%%%%%%%%%%%%%%%%%%%%%%%%%%%%%%%%%%%%%%%%%%%%%%%%%%%%%%%%%%

Consider a material representative volume element now defined in terms
of a periodic unit cell (PUC). Similar to the previous sections,
suppose that the PUC is subjected to boundary displacements
$\Delta\tenss{u}_0$ resulting in a uniform strain $\Delta\tenss{E}$
throughout the body. The strain and displacement fields in the PUC then admit the
following decomposition, recall also Eq.~\eqref{ec44}$_3$,
\begin{eqnarray}
\label{eq:disp}
\Delta\tenss{u}(\vek{x}) &=& \Delta\tenss{E} \cdot \vek{x} + \Delta\tenss{u}^*(\vek{x}),
\hspace{1cm}\forall\,\vek{x}\in\Omega,\;\;\tenss{u}=\tenss{u}_0\;
\forall\,\vek{x}\in S,\\ \label{eq:strain}
\Delta\tenss{\strain}(\vek{x}) &=& \Delta\tenss{E} + \Delta\tenss{\strain}^*(\vek{x}),
\hspace{1.65cm}\forall\,\vek{x}\in\Omega.
\end{eqnarray}
The first term in Eq.~(\ref{eq:disp}) corresponds to a displacement
field in an effective homogeneous medium with the same overall
properties as the composite. The fluctuation part $\tenss{u}^*$ enters
Eq.~(\ref{eq:disp}) as a consequence of the presence of
heterogeneities and has to disappear upon volume averaging,
see~\cite{Beran:1968:SCT} for further discussion.  This condition is
met for any periodic displacement field with the period equal to the
size of the unit cell under consideration,~\cite[and references
  therein]{Michel:1999:EPC}. The periodicity of $\tenss{u}^*$ further
implies that the average of $\tenss{\strain}^*$ in the unit cell
vanishes as well. Hence
\begin{equation}
\avg{\tenss{\strain}(\vek{x})} = \tenss{E} +
\avg{\tenss{\strain}^*(\vek{x})}, \hspace{1cm}\avg{\tenss{\strain}^*(\vek{x})} =
\frac{1}{\uc} \int_\uc \tenss{\strain}^*(\vek{x}) \de{\vek{x}} =
\tenss{0}.
\end{equation}

Derivation of the macroscopic response then relies on the application
of Hill's lemma~\cite{Hill:1963:EPT}. We now introduce a slight
inconsistency in the present formulation and assume that uniform
tractions $\Delta\tenss{p}_0$ (consistent with the macroscopic
stresses $\Delta\tenss{\Sigma},
\Delta\tenss{p}_0=\Delta\tenss{\Sigma}\cdot\tenss{n}, \tenss{n}$ being
the unit outward normal to the boundary of the PUC) rather than
displacements are prescribed to get
\begin{equation}\label{eq:hlemma5}
\left<\delta\tenss{\strain}\dcontr\Delta\tenss{\sigma}(\vek{x})\right> =
\delta\tenss{E}\dcontr\Delta\bmath{\Sigma}.
\end{equation}
Substituting the discretized form of the local strain increment
$\evek{\Delta\strain(\vek{x})}=\evek{\Delta{E}}+\emtrx{B}\evek{\Delta{r}^*}$
  together with local constitutive equation \eqref{eq:MT-1}$_1$ into
  Eq.~(\ref{eq:hlemma5}) gives
\begin{equation}\label{eq:femcreep}
\left[ 
\begin{array}{ll}
\displaystyle \frac{1}{\uc} \int_\uc \emtrx{L(\vek{x})}\de\uc &  
\displaystyle \frac{1}{\uc} \int_\uc \emtrx{L(\vek{x})}\emtrx{B}\de\uc \\
\displaystyle \frac{1}{\uc} \int_\uc \emtrx{B}\trn \emtrx{L(\vek{x})}\de\uc &
\displaystyle \frac{1}{\uc} \int_\uc \emtrx{B}\trn \emtrx{L(\vek{x})}\emtrx{B}\de\uc  
\end{array} 
\right]
\left\{ 
\begin{array}{c}
\evek{\Delta{E}} \\
\evek{\Delta{r}^*}
\end{array}
\right \} = 
\left\{ 
\begin{array}{c}
\displaystyle \evek{\Delta{\Sigma}} - \frac{1}{\uc} \int_\uc
\evek{\Delta\lambda}\de\uc \\
\displaystyle -\frac{1}{\uc} \int_\uc\emtrx{B}\trn\evek{\Delta\lambda}
\de\uc \\ 
\end{array}
\right\}.
\end{equation} 
Recall that in the framework of finite element method the matrix
$\emtrx{B}$ is often termed the geometrical matrix and the vector
$\evek{\Delta{r}^*}$ stores the increments of the fluctuation nodal
displacements. Finally, after rewriting the above equation as
\begin{equation}\label{eq:ve6} 
\left[ \begin{array}{cc}
\emtrx{K}_{11} & \emtrx{K}_{12}\\  
\emtrx{K}_{21} & \emtrx{K}_{22}\\
\end{array} \right]
\left\{ \begin{array}{c}
\evek{\Delta{E}}\\ \evek{\Delta{r}^*}
\end{array} \right\}= 
\left\{ \begin{array}{c}
\evek{\Delta{\Sigma}}+\evek{\Delta{F}^0}\\
\evek{\Delta{f}^0}\end{array} \right\}, 
\end{equation}  
and eliminating the fluctuating displacements vector
$\evek{\Delta{r}^*}$ we arrive at the incremental form of the
macroscopic constitutive law
\begin{equation}\label{eq:ve8}
\evek{\Delta{\Sigma}}=\emtrx{{L}^{\text{hom}}}\evek{\Delta{E}}+\evek{\Delta{\Lambda}},
\end{equation}
where
$$
\emtrx{{L}^{\text{hom}}}=\left(\emtrx{K}_{11}-\emtrx{K}_{12}\emtrx{K}_{22}^{-1}\emtrx{K}_{12}^
{\rm T}\right), \hspace{0.25cm} 
\evek{\Delta{\Lambda}}\;=\; -\evek{\Delta{F}^0}+
\left(\emtrx{K}_{12}\emtrx{K}_{22}^{-1}\evek{\Delta{f}^0}\right).$$
Also note that, when returning back to our original formulation with
the prescribed overall strain only, the system of equations
(\ref{eq:ve6}) reduces to
\begin{equation}\label{eq:ve9}
\emtrx{K}_{22}\evek{\Delta{r}^*}=-\emtrx{K}_{21}\evek{\Delta{E}}+\evek{\Delta{f}^0}.
\end{equation}

\subsubsection{Effective properties from binary images}\label{bbsec:MAM-results}  
%%%%%%%%%%%%%%%%%%%%%%%%%%%%%%%%%%%%%%%%%%%%%%%%%%%%%%%%%%%%%%%%%%%%%%

A simple example of a two-scale elastic homogenization, recall
Fig.~\ref{F:intro:2}, is presented here for the binary images plotted
in Figs.~\ref{F:rve:2} and~\ref{F:rve:3}. Both the aggregates and
mastic are assumed to be isotropic. The elastic modulus of the mastic
phase being equal to 1,600 MPa ($\nu_m=0.4$) is associated with the
value of the storage modulus at zero value of the loss modulus based
on the Cole Cole graph in Fig.~\ref{F:lab:1}(d). The elastic material
properties of aggregates are considered to be those of a spilite rock
with the elastic modulus equal to 60,000 MPa ($\nu_s=0.2$).

%%%%%%%%%%%%%%%%%%%%%
\begin{table}[ht]
\caption{Effective properties of mortar matrix}
\label{T:mortar}
\centering
\begin{tabular}{|l|c|c|c|c|c|c|}
\hline
{\small PUC} (Fig.~\ref{F:rve:3}) & (a) & (b) & (c) & (d) & FEM (d) & MT (d) \\
\hline 
$E^{\text{hom}}$ {\small [MPa]} & 2443 & 2670 & 2897 & 3150 & 3145 & 3462\\ 
$\nu^{\text{hom}}$ [-] & 0.38 & 0.38 & 0.38 & 0.38 & 0.38 & 38 \\
\hline
\end{tabular}
\end{table}
%%%%%%%%%%%%%%%%%%%%%

\begin{figure}[ht]
\begin{center}
\includegraphics[angle=0,width=8cm]{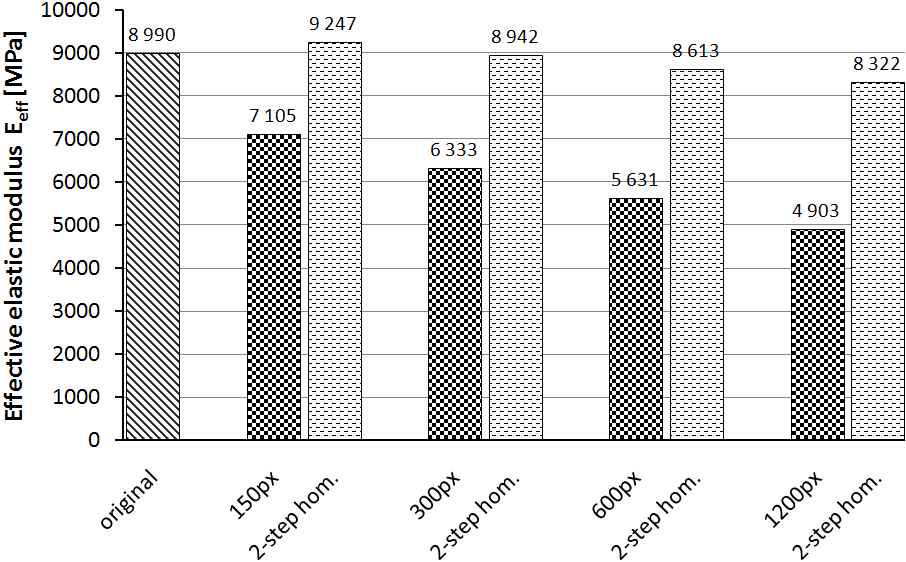}
\end{center}
\caption{Effective elastic modulus derived from images in Fig.~\ref{F:rve:2}}
\label{F:rve:5}
\end{figure}

Since dealing directly with the binary representation of real
microstructures the FFT method was adopted in this study. The
homogenized material properties on the mortar scale are listed in
Table~\ref{T:mortar}. Note that due to an hexagonal arrangement of
stones within the PUC the transverse material isotropy was generated.
In the second homogenization the effective properties of the mortar
are then used in place of the original mastic matrix when treating
individual RVEs in Fig.~\ref{F:rve:2}. The final results, promoting
the proposed two-step upscaling homogenization scheme, appear as
dash-pattern columns in Fig.~\ref{F:rve:5}. Also notice the
checkerboard columns, derived for the same microstructures but using
mastic material properties for the binder, clearly indicating a
considerable impact of eliminated stones on the predicted effective
properties.

%%%%%%%%%%%%%%%%%%%%%%%%%%%%%%%%%%%%%%%%%%%%%%%%%%%%%%%%%%%%%%%%%%%%%%%%%%%%%%%%%%%%%%%%%%%%%%
%%%%%%%%%%%%%%%%%%%%%%%%%%%%%%%%%%%%%%%%%%%%%%%%%%%%%%%%%%%%%%%%%%%%%%%%%%%%%%%%%%%%%%%%%%%%%%

\subsection{Statistically equivalent periodic unit cell}\label{bsec:sepuc}
%%%%%%%%%%%%%%%%%%%%%%%%%%%%%%%%%%%%%%%%%%%%%%%%%%%%%%%%%%%%%%%%%%%%%%
Although considerable savings in computational time can be achieved
with coarse RVEs, the analysis employing original microstructures
still presents a significant challenge particularly in view of a large
scale nonlinear analysis of multi-layered rodes carried out, possibly,
in the framework of multiscale FE$^2$
analysis~\cite{Fish:00:IJCSE,Kouznetsova:2002:MSCM}. Even at moderate
temperatures the stiffness of the binder might reduce by several
orders of magnitude resulting in a drastic mismatch between the
elastic moduli of binder and aggregates and consequently in a poor
convergence performance of the FFT method. Although some improvements
to the original formulation exist, they will not be considered in the
present study. Instead a rather different route is built taking
account of the so called statistically equivalent periodic unit
cell~\cite{Zeman:2001:EPG}.

\begin{figure}[ht]
\begin{center}
\begin{tabular}{c@{\hspace{10mm}}c@{\hspace{10mm}} c@{\hspace{10mm}}c}
\includegraphics[angle=0,width=2.6cm]{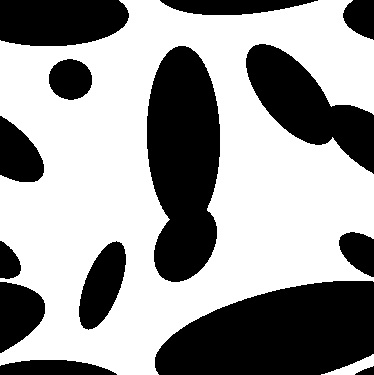}&
\includegraphics[angle=0,width=2.6cm]{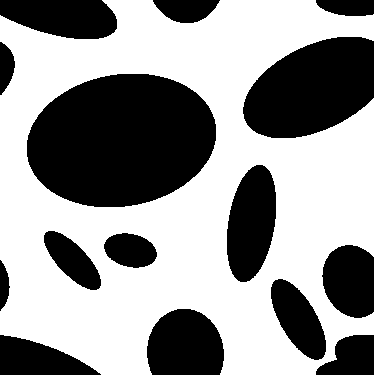}&
\includegraphics[angle=0,width=2.6cm]{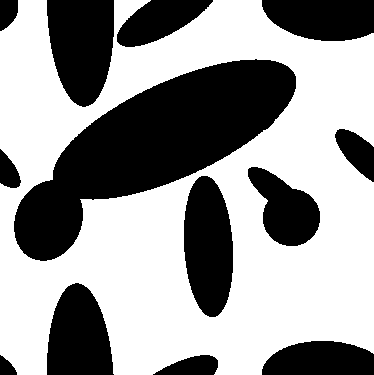}&
\includegraphics[angle=0,width=2.6cm]{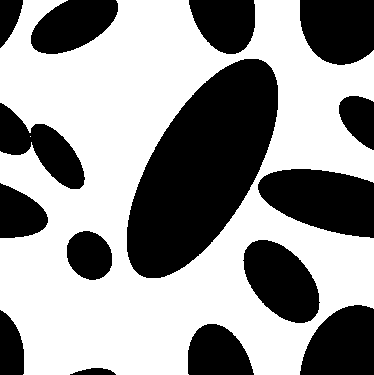}\\
(a) & (b) & (c) & (d)
\end{tabular}
\end{center}
\caption{Examples of SEPUCs corresponding to a binary image in Fig.~\ref{F:rve:2}(d):
(a) SEPUC 6,
(b) SEPUC 37,
(c) SEPUC 43,
(c) SEPUC 48}
\label{F:rve:6}
\end{figure}

In particular, suppose that the original microstructure can be
replaced by a certain artificial periodic unit cell that, from the
microstructure point of view, statistically resembles the real
material system in terms of, e.g. the two-point probability function,
see e.g.~\cite{Torquato:2002:RHM} for in depth discussion on various
statistical descriptors. Such a unit cell can be defined by the
following parameters: number of aggregates having elliptical shape,
size, position, orientation and aspect ratio of the axes of individual
ellipses. The size of stones is derived based on the cumulative
distribution function in Fig.~\ref{F:rve:1}. For example, if 10
stones are selected for a PUC then the smallest stone corresponds to
an average size of 10\% of the smallest stones determined from the
cumulative distribution function. The next stone then reflects the
size of the subsequent 10\% stones, etc. Examples of such unit cells
are depicted in Fig.~\ref{F:rve:6}.

%%%%%%%%%%%%%%%%%%%%%%%%%%%%%%%%%%%%%%%%%%%%%%%%%%%%%%%%%%%%%%%%%%%%%%%%%%%%%%%%%%%%%%%%%%%%%%
%%%%%%%%%%%%%%%%%%%%%%%%%%%%%%%%%%%%%%%%%%%%%%%%%%%%%%%%%%%%%%%%%%%%%%%%%%%%%%%%%%%%%%%%%%%%%%

These unit cells were derived by matching the two-point probability
function of the original microstructure, Fig.~\ref{F:rve:1}(d), and
the SEPUC. The underlying optimization problem was solved with the
help of the evolutionary algorithm GRADE
\cite{Ibrahimbegovic:2004,Kucerova:2007:PHD}. See also
~\cite{Zeman:2001:EPG,Matous:2000:GEI} for other applications of
genetic algorithm based solution strategies. It is worth mentioning
that no interpenetration constrain was introduced as it was naturally
enforced through the consistency of volume fractions of aggregates in
the SEPUC and targeted microstructure in Fig.~\ref{F:rve:1}(d).

%%%%%%%%%%%%%%%%%%%%%%%%%%%%%%%%%%%%%%%%%%%%%%%%%%%%%%%%%%%%%%%%%%%%%%%%%%%%%%%%%%%%%%%%%%%%%%
\begin{figure}[ht]
\begin{center}
\includegraphics[angle=0,width=8cm]{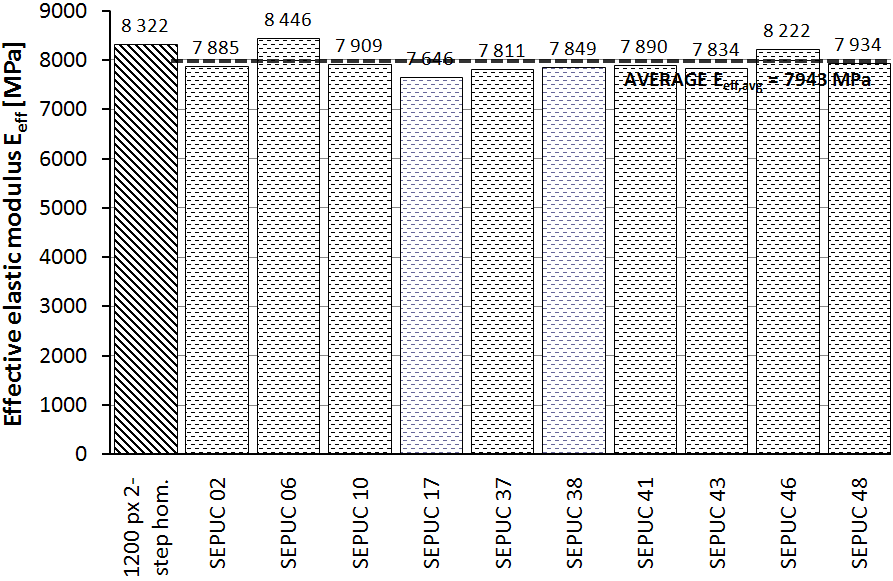}
\end{center}
\caption{Effective elastic modulus derived from SEPUCs in Fig.~\ref{F:rve:5}}
\label{F:rve:7}
\end{figure}
%%%%%%%%%%%%%%%%%%%%%%%%%%%%%%%%%%%%%%%%%%%%%%%%%%%%%%%%%%%%%%%%%%%%%%%%%%%%%%%%%%%%%%%%%%%%%%

As typical of genetic algorithms based optimization procedures, each
run results in a unique SEPUC with a slightly different arrangement of
stones, see Fig.~\ref{F:rve:6}. It therefore remains to confirm that
all cells provide the ``same'' macroscopic response. Note that for the
sake of efficiency the target microstructure in Fig.~\ref{F:rve:1}(d)
with mastic binder being replaced by mortar phase (the fourth column in
Table~\ref{T:mortar}) was used when generating artificial periodic
microstructures. The corresponding effective properties are
represented by the first column in Fig.~\ref{F:rve:7}. The remaining
columns refer to the effective elastic modulus for the selected SEPUCs
(100 such cells were generated). Although slightly different in their
geometrical details they all provide nearly the same macroscopic
response almost identical to the original microstructure. 

%%%%%%%%%%%%%%%%%%%%%
\begin{table}[ht]
\caption{Effective properties of asphalt}
\label{T:asphalt}
\centering
\begin{tabular}{|l|c|c|c|c|c|c|c|}
\hline
{\small SEPUC} (Fig.~\ref{F:rve:6}) & (a) & (b) & (c) & (d) & MT (d) & Fig.~\ref{F:rve:1}(d) & Fig.~\ref{F:intro:1}(c)\\
\hline 
$E^{\text{hom}}$ {\small [MPa]} & 8446 & 7811 & 7934 & 7834 & 7426 &  8322 & 8998\\ 
$\nu^{\text{hom}}$ [-] & 0.33 & 0.35 & 0.33 & 0.33 & 0.33 & 0.32 & 0.31\\
\hline
\end{tabular}
\end{table}
%%%%%%%%%%%%%%%%%%%%%

The Mori-Tanaka prediction found from the same upscaling procedure
assuming aggregates to be represented by statistically uniform
distribution of circular cylinders is also presented and compared in
Table~\ref{T:asphalt} with the selected estimates provided by SEPUCs
(first four columns), the reduced microstructure in
Fig.~\ref{F:rve:1}(d) (the last but one column) and the original
microstructure in Fig.~\ref{F:intro:1}(c) (last column). Although
still within an acceptable range the Mori-Tanaka method slightly
underestimates the macroscopic elastic response. This difference would
be even more pronounced if using mortar properties given by FFT
simulations.

In most practical applications, however, the mastic asphalt is
typically loaded beyond the elastic regime. The issue, whether the
geometrical invariance of SEPUCs outlasts even for a non-linear
response should be examined. This will be the principal objective 
of the next section.

\section{Generalized Leonov model for Mastic Asphalt mixture from two-step homogenization}\label{sec:LEONOV}
%%%%%%%%%%%%%%%%%%%%%%%%%%%%%%%%%%%%%%%%%%%%%%%%%%%%%%%%%%%%%%%%%%%%%%%%%%%%%%%%%%%%%%%%%%%%%%%%%%%%%%%%%%%%%%%%%%%%%%%%%%%%%
It has been experimentally observed~\cite{Shell-Bitumen} that both
asphalt binders and asphalt mixtures exhibit a considerable dependence
on the rate of loading and temperature. This phenomenon can be well
described by the Generalized Leonov (GL) model. Although some
experimental results point out a pressure dependent behavior of these
systems, the present study is limited to a Mises-type material with
negligible plastic volumetric deformation. The essential properties of
the GL model are presented in the next
section. Section~\ref{bsec:leonov-mastic} then summarizes the
experimental program to calibrate the GL model on mastic
scale. Upscaling towards macroscopic viscous properties of the GL
model is then addressed in
Sections~\ref{bsec:leonov-mortar}~-~\ref{bsec:leonov-MAM}.

\subsection{Basic properties and implementation of GL model}\label{bsec:leonov}
%%%%%%%%%%%%%%%%%%%%%%%%%%%%%%%%%%%%%%%%%%%%%%%%%%%%%%%%%%%%%%%%%%%%%%
Combing the Eyring flow model for the plastic component of the shear
strain rate
\begin{equation}
\frac{\de e_p}{\de t} = \frac{ 1 }{ 2A } \sinh \frac{ \tau }{ \tau_0 },
\label{eq:Eyring}
\end{equation} 
with the elastic shear strain rate $\de e_e/\de t$ yields the
one-dimensional Leonov constitutive model~\cite{Leonov}
\begin{equation}
\frac{\de e}{\de t}  = \frac{\de e_e}{\de t}  + \frac{\de e_p}{\de t}
= \frac{\de e_e}{\de t} + \frac{\tau}{\eta({\de e_p/\de t})},
\label{eq:NMaxwell}
\end{equation}
where the shear-dependent viscosity $\eta$ is provided by
\begin{equation}
\eta({\de e_p/\de t}) = \frac{ \eta_0 \tau }{ \tau_0 \sinh( \tau /
\tau_0 ) } = \eta_0 a_\sigma( \tau ).
\label{eq:Sleonov}
\end{equation}
In Eq.~(\ref{eq:Eyring}), $A$ and $\tau_0$ are material parameters;
$a_\sigma$ that appears in Eq.~(\ref{eq:Sleonov}) is the stress shift
function with respect to the zero shear viscosity $\eta_0$~(viscosity
corresponding to a viscoelastic response). Clearly, the phenomenological
representation of Eq.~(\ref{eq:NMaxwell}) is the Maxwell model with
the variable viscosity $\eta$. Experimental evidence proving analogy
between stress and temperature allows us to extend
Eq.~\eqref{eq:Sleonov} by adding a temperature ($T$) dependent shift
factor $a_T$ to get
\begin{equation}
\eta({\de e_p/\de t}) = \eta_0 a_\sigma( \tau )a_T(T).
\label{eq:eta-sig-T}
\end{equation}

To describe multi-dimensional behavior of the material, the
generalized compressible Leonov model, equivalent to the generalized
Maxwell chain model, can be used~\cite{Tervoort:Hab}. The viscosity
term corresponding to the $\mu$-th unit receives the form
\begin{equation}
\eta_\mu = \eta_{0,\mu} a_\sigma( \tau_{\mathrm{eq}} )a_T(T),
\end{equation}
where the equivalent shear stress $\tau_{\mathrm{eq}}$ is provided by
\begin{equation}
\tau_{\mathrm{eq}} = \sqrt{\frac{1}{2} \tenss{s}\dcontr\tenss{s}},
\end{equation}
and $\tenss{s}$ is the deviatoric stress tensor. Admitting only
small strains and isotropic material, a set of constitutive equations
defining the generalized compressible Leonov model can be written as
\begin{eqnarray}
\stress_m & = & K \strain_v, \label{eq:GLeonov1} \\
\frac{\de \tenss{s} }{ \de t } & = & \sum_{\mu=1}^{M} 2G_\mu \left(
\frac{\de \tenss{e}}{\de t} - \frac{\de \tenss{e}_{p,\mu}}{\de t} \right),
\hspace{0.5cm}\tenss{s} \,= \, \sum_{\mu = 1}^{M} \tenss{s}_\mu,
\\  \tenss{s}_\mu & = & 2 \eta_\mu \frac{\de \tenss{e}_{p,\mu}}{\de t} = 2
\eta_{0,\mu} a_\sigma( \tau_{\mathrm{eq}} )a_T(T)\frac{\de \tenss{e}_{p,\mu}}{\de 
t},
\label{eq:GLeonov2}
\end{eqnarray}
where $\stress_m$ is the mean stress, $\strain_v$ is the volumetric
strain, $K$ is the bulk modulus and $G_\mu$ is the shear modulus of
the $\mu$-th unit.

While a fully implicit scheme has been implemented
in~\cite{Sejnoha:MCE:2004} to integrate the system of
Eqs.~\eqref{eq:GLeonov1}~-~\eqref{eq:GLeonov2}, a fully explicit Euler
forward (EF) integration scheme is adopted here for simplicity. Thus
to avoid numerical instabilities a relatively short time step must be
prescribed. This is merely attributed to the fact that viscous
parameters vary nonlinearly with stress, but in the context of EF
scheme are assumed constant over the time step.

Provided that the total strain rate is constant
during integration a new state of stress in the matrix phase at the
end of the current time step assumes the form 
%(subscript $\mathsf{m}$, identifying the matrix phase, is dropped
%from subsequent equations)
\begin{eqnarray}
\sigma_m( t_i ) &=& \sigma_m( t_{i-1} ) + K\Delta\strain_v, \label{eq:smean} \\
\evek{s(\scal{t_i})} &=& \evek{s(\scal{t_{i-1}})}
+ 2\widehat{G}(t_{i-1})\emtrx{Q}\evek{\Delta e} + \evek{\Delta \estress( \scal{t_{i-1}} )},
\label{eq:sti}
\end{eqnarray}
where $t_i$ is the current time at the end of the $i$-th time
increment; $\sigma_m$ is the elastic mean stress, $\evek{s}$ stores
the deviatoric part of the stress vector $\evek{\sigma}$ and $\evek{
  \Delta e}$ is the deviatoric part of the total strain
increment. Assuming the shear relaxation modulus $\widehat{G}(t)$ can
be represented by a Dirichlet series expansion in the form
\begin{equation}
\widehat{G}(t)=\sum_{\mu=1}^{M}G_\mu\text{exp}\left(-\frac{t}{\theta_{\mu}a_{\sigma}a_T}\right),\label{eq:relaxation}
\end{equation}
and with reference to the
EF integration scheme the time dependent variables at time
instant $t_{i-1}$ receive the form
\begin{eqnarray}
\widehat{G} &=&
\sum_{\mu=1}^{M} {G}_\mu\frac{ \theta_\mu a_\sigma( t_{i-1} )a_T( t_{i-1} )}{  \Delta t} \left[ 1 -
\exp\left( -\frac{ \Delta t}{ \theta_\mu a_\sigma( t_{i-1} )a_T( t_{i-1} )}\right)\right],
\label{ea:Gve}
\\
%\evek{\Delta \estress(\scal{t_{i-1}}}) &=&
\evek{\Delta \estress} &=&
- \sum_{\mu=1}^{M} \left[  1 - \exp\left( -\frac{ \Delta t}
	{\theta_\mu a_\sigma( t_{i-1} )a_T( t_{i-1} )} \right)\right]\evek{s_\mu( \scal{t_{i-1}} )},
\label{eq:deti}
\end{eqnarray}
where $G_\mu$ represents the elastic shear modulus in the $\mu$-th
unit of the Maxwell chain model, $\theta_\mu$ is the associated
relaxation time, $\evek{s_{\mu}\scal(t_{i-1})}, \mu=1,2,\ldots,M,$ is
the deviatoric stress vector in individual units evaluated at the
beginning of the new time increment $\Delta{t} = t_i - t_{i-1}$, and
$M$ is the assumed number of Maxwell units in the chain model;
$a_\sigma(t_{i-1})$ and $a_T(t_{i-1})$ are the stress and temperature
shift factors, respectively. The relaxation shear moduli $G_\mu$ are
found from a Dirichlet series expansion of creep compliance function
presented, e.g. in Fig.~\ref{F:lab:1}(c) by invoking the
correspondence principle. 
An example of a direct application of this principle is given in Appendix.

According to Eq.~\eqref{eq:Sleonov} the
stress shift factor $a_\sigma$ reads
\begin{eqnarray}
a_\sigma(t_{i-1}) & = &
\frac{\tau_{eq}(t_{i-1})}{\tau_0} / \sinh\frac{\tau_{eq}(t_{i-1})}{\tau_0}, 
\label{eq:asogti}
\end{eqnarray}
where the equivalent stress $\tau_{eq}(t_{i-1})$ follows from
\begin{eqnarray}
\tau_{eq}(t_{i-1}) &=& \sqrt{\frac{1}{2}\evek{s( \scal{t_{i-1}} )}\trn\emtrx{Q}^{-1}\evek{s( \scal{t_{i-1}} )}},
\label{eq:tauti-1}
\end{eqnarray}
and
\begin{eqnarray}
\emtrx{Q} & = & \mbox{diag}
\left[1,1,1,\frac{1}{2},\frac{1}{2},\frac{1}{2}\right].
\label{eq:Q}
\end{eqnarray}
The temperature shift factor is fitted to Williams-Landel-Ferry (WLF) equation written as
\begin{eqnarray}
a_T = \text{exp}\left(\frac{-C_1(T-T_0)}{C_2+(T-T_0)}\right),
\label{eq:at}
\end{eqnarray}
where $C_1, C_2$ are model parameters, $T_0$ is the reference temperature
and $T$ is the actual temperature.

\subsection{Rate and temperature dependent behavior of mastic from laboratory experiments}\label{bsec:leonov-mastic}
%%%%%%%%%%%%%%%%%%%%%%%%%%%%%%%%%%%%%%%%%%%%%%%%%%%%%%%%%%%%%%%%%%%%%%

Referring to~\cite{Valenta:2004:Madeira} we anticipated to derive the
model parameter $\tau_0$, needed in turn for the determination of the
shift factor $a_\sigma$, from a set of experiments when loading
individual specimens in shear at constant strain rate. The available
experimental devices, however, failed to provide the required viscous
flow at zero elastic strain increment~\cite{Tervoort:Hab}. Additional
difficulties in controlling the applied normal pressure when twisting
the specimens immediately promoted the following alternative approach:

\begin{itemize}
\item Construction of the compliance master curve from the
  measurements of dynamic moduli at a selected range of frequencies and
  temperatures to first get the parameters of Eq.~\eqref{eq:at} approximating
  the temperature shift factor $a_T$.
\item Determination of the model parameter $\tau_0$ in
  Eq.~\eqref{eq:asogti} by horizontally shifting the experimentally
  derived compliance functions obtained from creep measurements for a
  given temperature at various stress levels.
\end{itemize}

The experimental program for the determination of frequency
characteristics of the mastic phase was carried out jointly at the
Central laboratory of Eurovia Services, Ltd. and the department of
Road Structures at the Faculty of Civil Engineering, Czech Technical
University in Prague. Standard dynamic shear rheometer (DSR) tests
were conducted under cyclic loading undergoing a temperature and
frequency sweep from 0 to 80$^0$C and 0.1 to 100 Hz, respectively.
For low temperature tests (from 0 to 20$^0$C) a mastic film was placed
between two plates (a lower fixed and an upper oscillating) of
diameter $d=8$ mm and pressed to maintain a constant height $h=2$ mm
during oscillations. For high temperatures (from 30$^0$ to 80$^0$C) the
$d/h$ ratio equal to 25/1 mm was used. The examined mastic consisted of
37$\%$ of limestone filler and 62,5$\%$ of polybitumen AMe 65
asphalt with bulk density $\rho_m$=1025 kgm$^{-3}$. The filler
aggregates were in the following grading range: 2 mm sieve - 100$\%$,
0.125 mm sieve - 81.1$\%$, 0.063 mm sieve - 70.3$\%$, with $\rho_s$=2730
kgm$^{-3}$.

The storage ($G^{'}$) and loss ($G^{''}$) dynamic moduli were found by
loading a mastic specimen in shear under strain controlled regime
\begin{equation}
\gamma=\overline{\gamma}\text{sin}(t\omega),
\hspace{5mm}\text{resulting in}\hspace{5mm}
\tau=\overline{\tau}\text{sin}(t\omega+\delta),
\end{equation}
where $\overline{\gamma}$ and $\omega$ are the prescribed shear strain
and angular frequency, respectively, and $\delta$ represents the phase
shift between stress and strain. The corresponding moduli then follow from
\begin{equation}
G^{'}=\frac{\overline{\tau}}{\overline{\gamma}}\text{cos}(\delta),
\hspace{1cm}
G^{''}=\frac{\overline{\tau}}{\overline{\gamma}}\text{sin}(\delta).
\end{equation}
The frequency sweep was selected such as to keep the resulting
stresses within the viscoelastic limit.

%%%%%%%%%%%%%%%%%%%%%%%%%%%%%%%%%%%%%%%%%%%%%%%%%%%%%%%%%%%%%%%%%%%%%%%%%%%%%%%%%%%%%%%%%%%%%%
\begin{figure}[ht]
\begin{center}
\begin{tabular}{c@{\hspace{1mm}}c}
\includegraphics[angle=0,width=6.75cm]{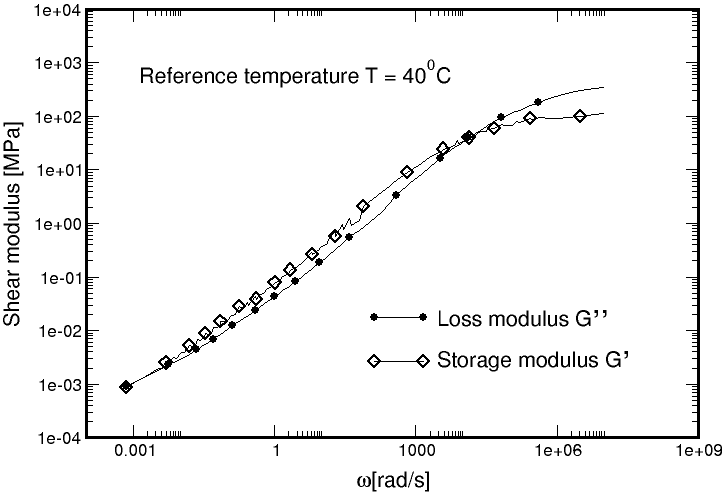}&
\includegraphics[angle=0,width=6.75cm]{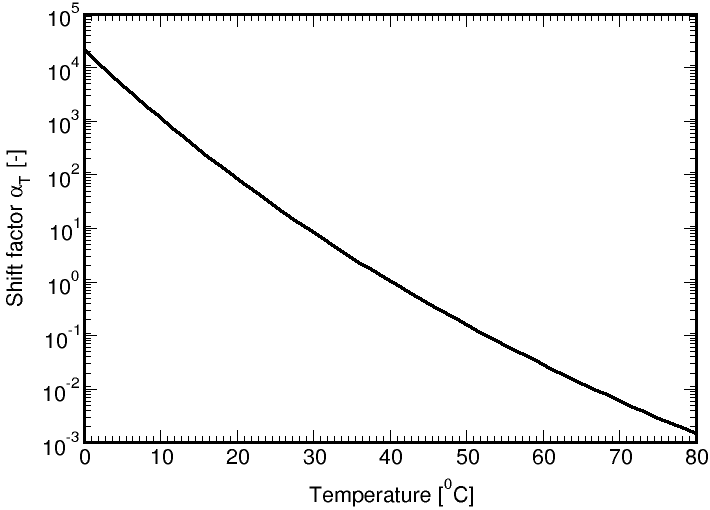}\\
(a)&(b)\\
\includegraphics[angle=0,width=6.75cm]{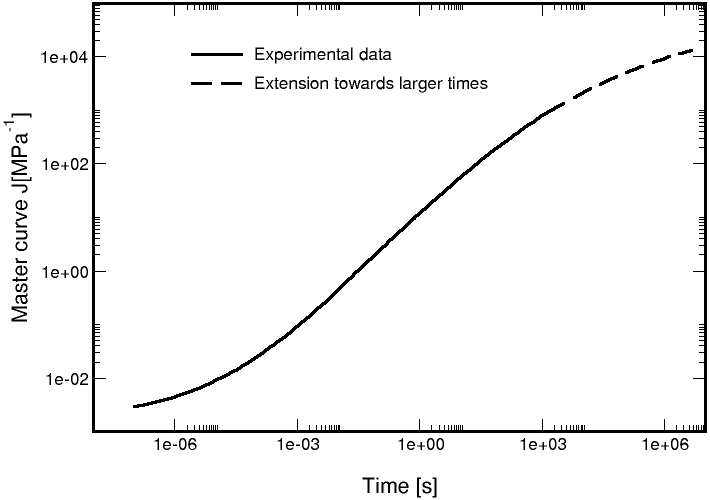}&
\includegraphics[angle=0,width=6.75cm]{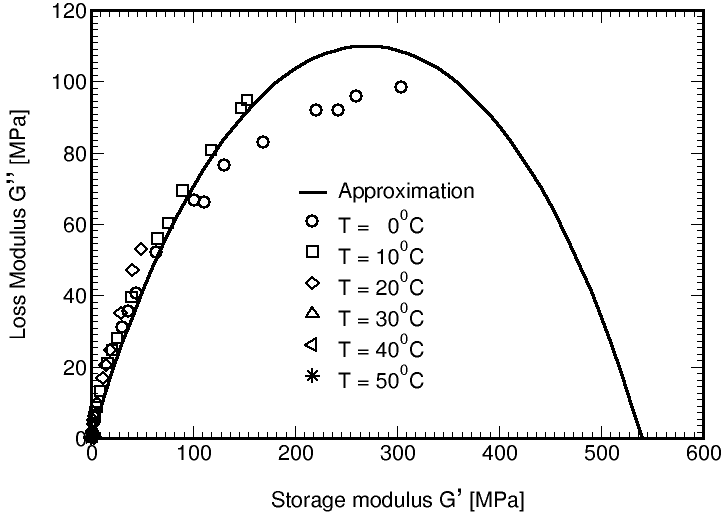}\\
(c)&(d)
\end{tabular}
\end{center}
\caption{(a) Dynamic moduli, (b) Shift factor, (c) Master curve, (d) Cole Cole graph}
\label{F:lab:1}
\end{figure}
%%%%%%%%%%%%%%%%%%%%%%%%%%%%%%%%%%%%%%%%%%%%%%%%%%%%%%%%%%%%%%%%%%%%%%%%%%%%%%%%%%%%%%%%%%%%%%

The master curves of dynamic shear moduli were constructed by
superimposing the dynamic data measured at different temperatures for
a given range of frequencies. Accepting the temperature superposition
principle the resulting master curves seen in Fig.~\ref{F:lab:1}(a)
were simply derived by horizontally shifting individual curves to
comply with the data one would obtain for the 40$^0$C reference
temperature at much longer (data for temperatures above 40$^0$C) or
shorter (data for temperatures below 40$^0$C) times if performing
standard creep measurements. The dynamic moduli master curves were
subsequently converted into a time-dependent compliance function at
40$^0$C employing the Fourier transform. The result appears in
Fig.~\ref{F:lab:1}(c). For the available temperature range the
resulting master curve can reliably cover the viscoelastic creep
response over the time span up to 1000 s. To enhance numerical
stability this curve was artificially prolonged by two additional
decades, see the dashed line in Fig.~\ref{F:lab:1}(c). The associated
temperature dependent variation of shift factor $a_T$ fitted to
Eq.~\eqref{eq:at} is plotted in Fig.~\ref{F:lab:1}(b). The
corresponding parameters are stored in Table~\ref{T:shift}. Finally,
The experimentally-obtained values of $G^{'}$ and $G^{''}$ are plotted
in the so-called Cole-Cole diagram in Fig.~\ref{F:lab:1}(d) used in
the present study to estimate the elastic modulus of mastic phase,
recall Section~\ref{bbsec:MAM-results}. Note that the master curves in
Figs.~\ref{F:lab:1}(a),(c) were provided directly by a built in
software of DSR devices.

\begin{figure}[ht]
\begin{center}
\begin{tabular}{c@{\hspace{1mm}}c}
\includegraphics[angle=0,width=6.75cm]{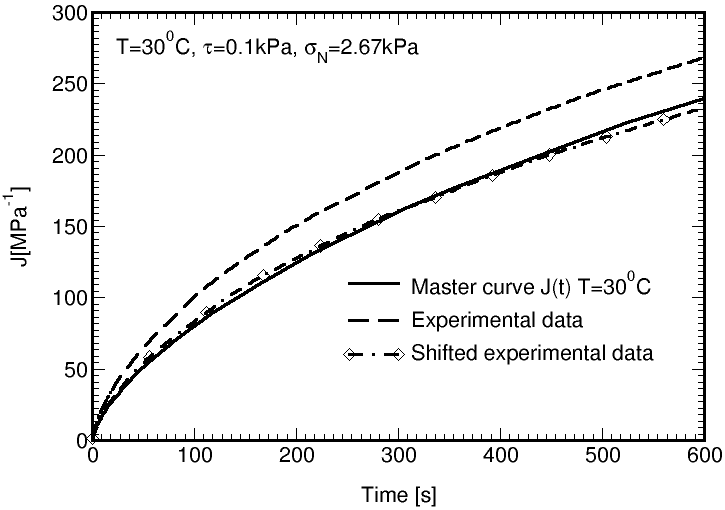}&
\includegraphics[angle=0,width=6.75cm]{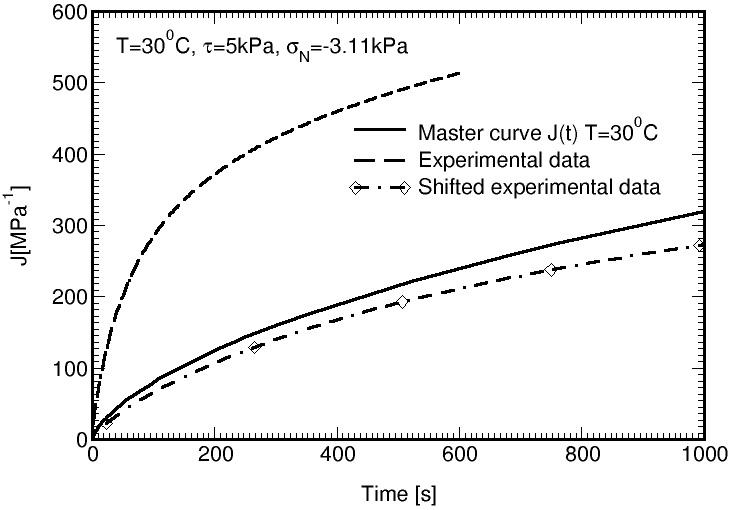}\\
(a)&(b)
\end{tabular}
\end{center}
\caption{Model performance - effect of the applied stress level : (a) $\tau=100$Pa, (b) $\tau=5000$Pa}
\label{F:lab:2}
\end{figure}

The second class of experiments delivered the creep data at constant
temperature $T=30^0$C but at different magnitudes of the applied
load. These data together with the previous results enabled us to
obtain the necessary parameter $\tau_0$.  To that end, the original
master curve in Fig.~\ref{F:lab:1}(c) was first converted to comply
with the one for 30$^0$C. A horizontal shift of four available creep
compliance functions finally served to estimate the parameter
$\tau_0$, see Table~\ref{T:shift}. Additional creep experiments
performed at temperature 50$^0$C were used to validate not only the
estimated value of $\tau_0$ but also the temperature-stress
superposition principle expressed, with reference to so called reduced
time, as
\begin{equation}
\varphi_t=\int_0^t\frac{\de{t}}{a_T[T(t)]\times a_\sigma[\tau_{eq}(t)]}.
\end{equation}
The results shown in Fig.~\ref{F:lab:2} confirm applicability of the
GL model to well represent the temperature and rate dependent behavior
of mastic material.

%%%%%%%%%%%%%%%%%%%%%
\begin{table}[ht]
\caption{Model parameters of Eqs.~\eqref{eq:Sleonov} and~\eqref{eq:at}}
\label{T:shift}
\centering
\begin{tabular}{|c|c|c|c|c|}
\hline
Scale & $T_0$  &$C_1$& $C_2$& $\tau_0$ [Pa]\\ 
\hline 
Mastic & 40$^0$C & 38.56 & 195.42 & 1150\\
Mortar & 40$^0$C & 38.56 & 195.42 & 2600\\
MAm & 40$^0$C & 38.56 & 195.42 & 2951\\
\hline
\end{tabular}
\end{table}
%%%%%%%%%%%%%%%%%%%%%

\begin{figure}[ht]
\begin{center}
\begin{tabular}{c@{\hspace{1mm}}c}
\includegraphics[angle=0,width=6.75cm]{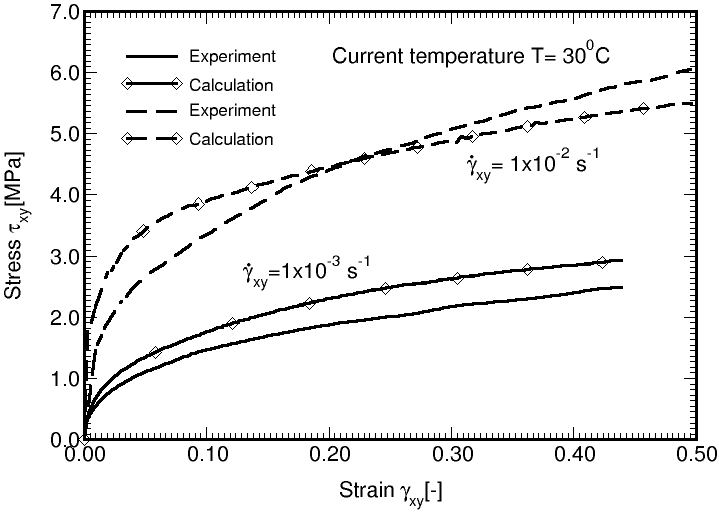}&
\includegraphics[angle=0,width=6.75cm]{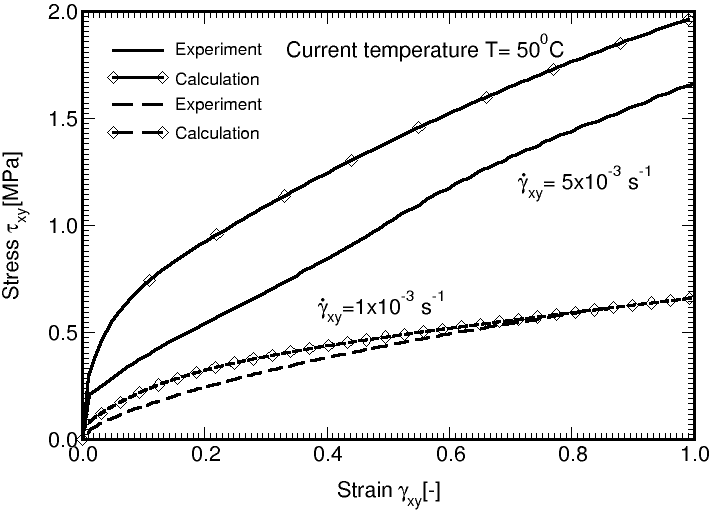}\\
(a)&(b)
\end{tabular}
\end{center}
\caption{Tensile loading at constant strain rate - experimental measurements vs. numerical predictions: (a) $T=30^0$C, (b) $T=50^0$C}
\label{F:lab:3}
\end{figure}

Further support for the use of GL model can be gained from the results
presented in Fig.~\ref{F:lab:3}. Therein, the numerical predictions of
static behavior at different strain rates and temperatures are
compared with experimentally obtained data. Note that this set of
experiments was not considered to calibrate the material
model. Realizing a difficulty of performing reproducible tests with
such a material suggests almost a ``remarkable'' match.  

%%%%%%%%%%%%%%%%%%%%%%%%%%%%%%%%%%%%%%%%%%%%%%%%%%%%%%%%%%%%%%%%%%%%%%%%%%%%%%%%%%%%%%%%%%%%%%
%%%%%%%%%%%%%%%%%%%%%%%%%%%%%%%%%%%%%%%%%%%%%%%%%%%%%%%%%%%%%%%%%%%%%%%%%%%%%%%%%%%%%%%%%%%%%%

%%%%%%%%%%%%%%%%%%%%%%%%%%%%%%%%%%%%%%%%%%%%%%%%%%%%%%%%%%%%%%%%%%%%%%%%%%%%%%%%%%%%%%%%%%%%%%
%%%%%%%%%%%%%%%%%%%%%%%%%%%%%%%%%%%%%%%%%%%%%%%%%%%%%%%%%%%%%%%%%%%%%%%%%%%%%%%%%%%%%%%%%%%%%%

\subsection{Response of mortar from virtual experiments}\label{bsec:leonov-mortar}
%%%%%%%%%%%%%%%%%%%%%%%%%%%%%%%%%%%%%%%%%%%%%%%%%%%%%%%%%%%%%%%%%%%%%%
Following on the heels of experimental work discussed in the previous
section, a virtual set of experiments is proposed to arrive at a
homogenized master curve and associated temperature and stress
dependent shift factors on the scale of mortar. To be consistent with
Section~\ref{bsec:sepuc} the PUC in Fig.~\ref{F:rve:3}(d) is chosen to
represent an RVE for numerical analysis. The concept of first order
homogenization of periodic fields outlined in Section~\ref{bbsec:FEM}
is given the preference to deliver the homogenized creep response of
the mortar phase at different temperature and stress levels.

\begin{figure}%[ht]
\begin{center}
\begin{tabular}{c@{\hspace{1mm}}c}
\includegraphics[angle=0,width=6.75cm]{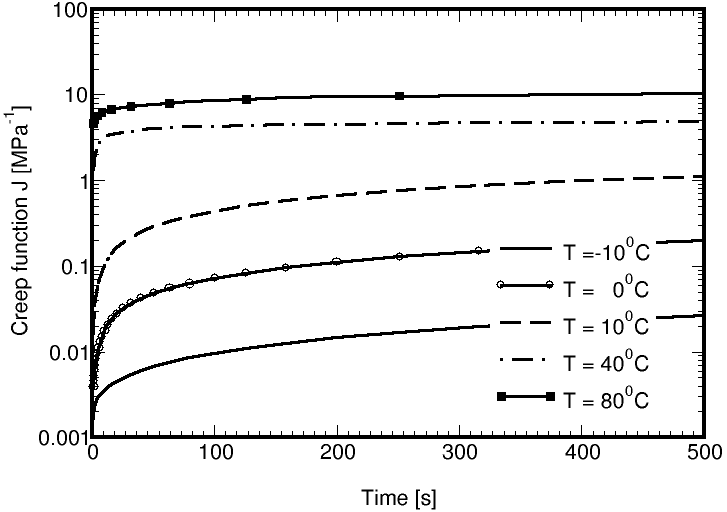}&
\includegraphics[angle=0,width=6.75cm]{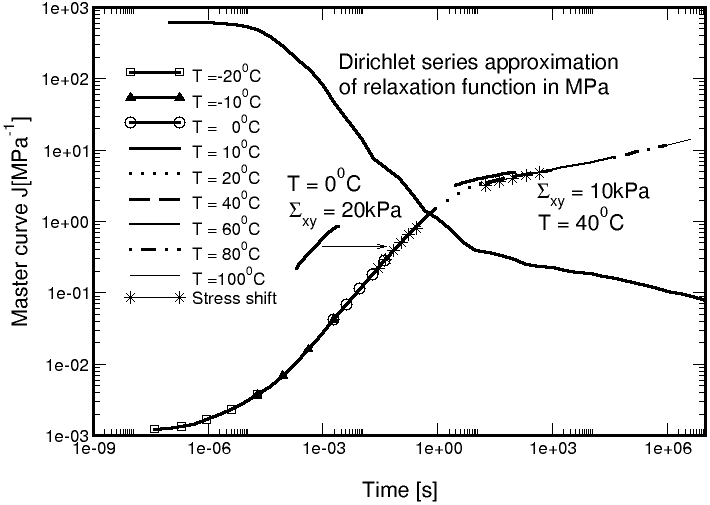}\\
(a)&(b)
\end{tabular}
\end{center}
\caption{(a) Creep data at different temperatures for reference stress $\Sigma_{xy}=1$kPa, (b) Master curve for reference temperature $T=40^0$C}
\label{F:PHA:1}
\end{figure}

First, a uniformly distributed range of temperatures from $0^0$C to
$100^0$C was considered to provide for a temperature dependent
viscoelastic behavior of mortar loaded in shear by the remote stress
$\Sigma_{yx}=1$kPa. Individual curves plotted in Fig.~\ref{F:PHA:1}(a)
were then horizontally shifted to give the homogenized creep
compliance master curve seen in Fig.~\ref{F:PHA:1}(b) and consequently
estimates of the parameters of WLF equation~\eqref{eq:at}. The
parameters $C_1, C_2$ for the reference temperature $T=40^0$C are
stored in Table.~\ref{T:shift} surprisingly suggesting no differences
between individual scales. The stiffening, observed for high
temperatures, is attributed to a volumetric locking owing to a very
low shear modulus approaching to zero. This in turn yields the Poisson
ratio close to 0.5 in a finite zone of the binder phase that is
progressed over the entire unit cell as seen in
Fig.~\ref{F:PHA:1}(b). Note that such a microstructure dependent
response can hardly be represented by the Mori-Tanaka method, see
appendix for further illustration.

The second set of creep experiments was conducted at two different
temperatures and two different levels of the remote stress
$\Sigma_{xy}$.  The two representative results, [$40^0$C, 10kPa] and
[$0^0$C, 20kPa], are plotted as solid lines in Fig.~\ref{F:PHA:1}(b).
The indicated horizontal shifts identified with the corresponding
star-lines then supplement the necessary data for the derivation of
stress dependent shift factor $a_\sigma$ fitted to
Eq.~\eqref{eq:asogti}. The searched parameter $\tau_0$ is available in
Table.~\ref{T:shift}. Note that prior to shifting the curves the
result derived for $0^0$C was thermally adjusted to be consistent with
the $40^0$C master curve.

%%%%%%%%%%%%%%%%%%%%%%%%%%%%%%%%%%%%%%%%%%%%%%%%%%%%%%%%%%%%%%%%%%%%%%%%%%%%%%%%%%%%%%%%%%%%%%
%%%%%%%%%%%%%%%%%%%%%%%%%%%%%%%%%%%%%%%%%%%%%%%%%%%%%%%%%%%%%%%%%%%%%%%%%%%%%%%%%%%%%%%%%%%%%%

\subsection{Response of MAm from virtual experiments}\label{bsec:leonov-MAM}
%%%%%%%%%%%%%%%%%%%%%%%%%%%%%%%%%%%%%%%%%%%%%%%%%%%%%%%%%%%%%%%%%%%%%%
Derivation of the macroscopic creep compliance master curve for a
Mastic Asphalt mixture follows the general scheme sketched in the
previous section. A statistically equivalent periodic unit cell
introduced in Section~\ref{bsec:sepuc} is considered for numerical
simulations. The principal assumption of restricting the material
model on every scale to one particular format is brought to light here
by suggesting the binder be well represented by the homogenized GL
model predicted on the mortar scale.

However, for problems involving significant nonlinearities, the
fundamental statement borrowed from elasticity solutions that any
statistically equivalent periodic unit cell is equally suitable should
be scrutinized first.  

%%%%%%%%%%%%%%%%%%%%%%%%%%%%%%%%%%%%%%%%%%%%%%%%%%%%%%%%%%%%%%%%%%%%%%%%%%%%%%%%%%%%%%%%%%%%%%
\begin{figure}[ht]
\begin{center}
\begin{tabular}{c@{\hspace{1mm}}c}
\includegraphics[angle=0,width=6.65cm]{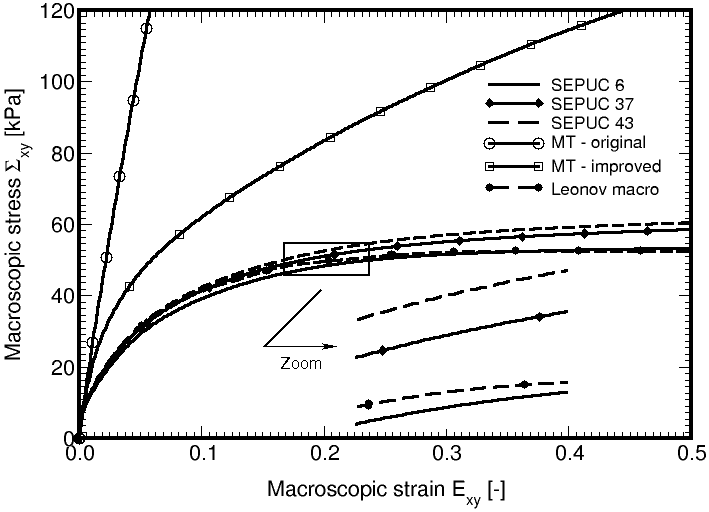}&
\includegraphics[angle=0,width=6.75cm]{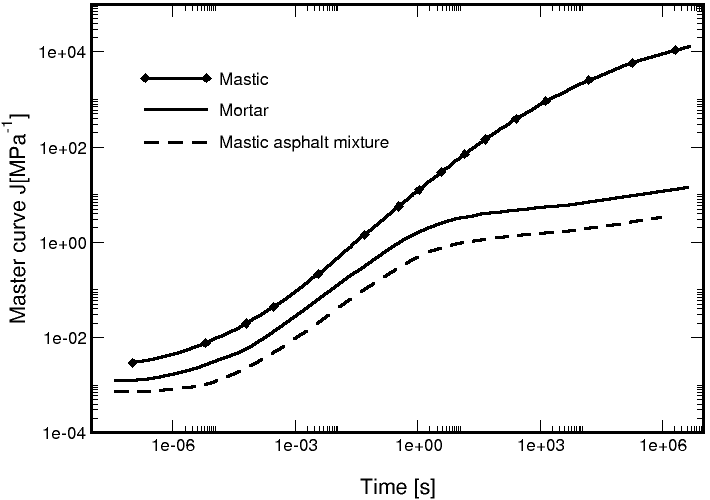}\\
(a)&(b)
\end{tabular}
\end{center}
\caption{(a) Macroscopic response for various SEPUCs $T=20^0$C, $\dot{E}_{xy}=10^{-4}$, (b) Master curves on individual scales for reference temperature $T=40^0$C}
\label{F:MAM:1}
\end{figure}
%%%%%%%%%%%%%%%%%%%%%%%%%%%%%%%%%%%%%%%%%%%%%%%%%%%%%%%%%%%%%%%%%%%%%%%%%%%%%%%%%%%%%%%%%%%%%%

\begin{figure}[ht]
\begin{center}
\begin{tabular}{c@{\hspace{5mm}}c@{\hspace{5mm}}c@{\hspace{5mm}}c}
\includegraphics[angle=0,width=3.0cm]{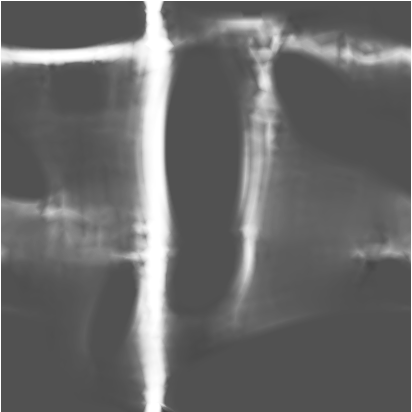}&
\includegraphics[angle=0,width=3.0cm]{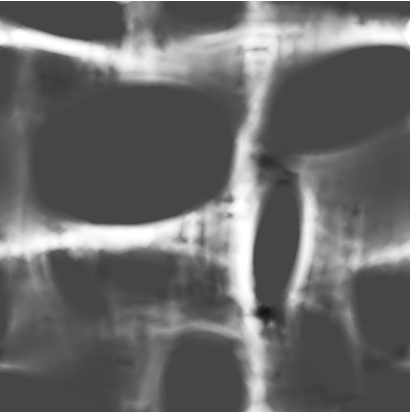}&
\includegraphics[angle=0,width=3.0cm]{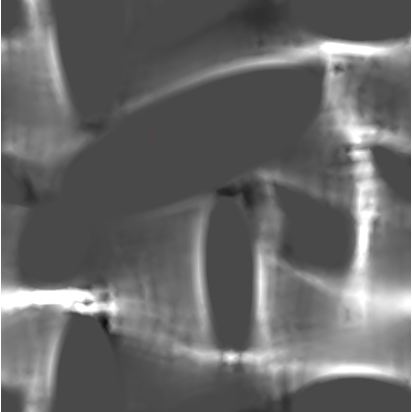}&
\includegraphics[angle=0,width=3.0cm]{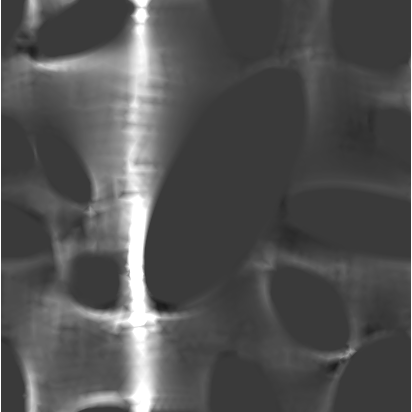}\\
(a)&(b)&(c)&(d)
\end{tabular}
\end{center}
\caption{Distribution of local shear strain:
(a) SEPUC 6,
(b) SEPUC 37,
(c) SEPUC 43,
(d) SEPUC 48}
\label{F:MAM:2}
\end{figure}

To address this issue the statistically equivalent unit cells in
Fig.~\ref{F:rve:6} were subjected to a remote shear strain rate
$\dot{E}_{xy}=10^{-4}$. The resulting homogenized stress-strain curves
for temperature $T=20^0$C are shown in Fig.~\ref{F:MAM:1}(a). Although
no ``perfect match'' is observed, the difference in estimated load
bearing capacities is in the same range as the predicted elastic
constants, see e.g. Table~\ref{T:asphalt}, not exceeding 10$\%$.  On
the contrary, the distribution of local fields varies considerably as
seen in Fig.~\ref{F:MAM:2}. While a highly localized distribution of
shear strain $\gamma_{xy}^m$ in the mortar phase, identified with the
lowest bearing capacity in Fig.~\ref{F:MAM:1}(a), is evident in
Fig.~\ref{F:MAM:2}(a), the variation of this quantity in
Fig.~\ref{F:MAM:2}(c) shows a rather distributed character
consequently resulting in a slightly stiffer response on the
macroscale.

Although certainly more accurate, the detailed finite element
simulations are in general computationally very expensive and often
call for less demanding alternatives such as the Mori-Tanaka method.
Unfortunately, the corresponding results also plotted in
Fig.~\ref{F:MAM:1}(a) clearly expose essential limitations of
two-point averaging schemes, unable to capture localization phenomena
observed in composites with a highly nonlinear response of the binder
phase~\cite{Sejnoha:MCE:2004}. On that ground, a considerably more
refined variant of the TFA method would be
necessary~\cite{Dvorak:1994:ITFA,Chaboche:2001:TMBIDM}. In this paper,
however, the attention is limited two a two-phase composite only as
discussed in Section~\ref{bbsec:MT}. The presented results in
Fig.~\ref{F:MAM:1}(a) correspond to a classical formulation with the
localization and transformation tensors calculated only once being
functions of elastic properties of individual constituents
(MT-original), and to the formulation where these tensors are updated
after each time step taking into account increasing compliance of the
mortar phase with time (MT-improved). Note that even the latter case
produces much stiffer response, although at a fraction of time, when
compared to the FE results. 

%%%%%%%%%%%%%%%%%%%%%%%%%%%%%%%%%%%%%%%%%%%%%%%%%%%%%%%%%%%%%%%%%%%%%%%%%%%%%%%%%%%%%%%%%%%%%%
%%%%%%%%%%%%%%%%%%%%%%%%%%%%%%%%%%%%%%%%%%%%%%%%%%%%%%%%%%%%%%%%%%%%%%%%%%%%%%%%%%%%%%%%%%%%%%

Regarding the ``similarity'' of macroscopic response from various
SEPUCs, the SEPUC No. 37 was selected to provide data needed in the
calibration of the macroscopic GL model. Virtual numerical tests
identical to those in Section~\ref{bsec:leonov-mortar} were again
performed to give first the homogenized master curve displayed in
Fig.~\ref{F:MAM:1}(b) and consequently the model parameters of
Eqs.~\eqref{eq:asogti} and~\eqref{eq:at} available in
Table~\ref{T:shift}. Note again the same temperature shift as obtained
already for lower scales. This is evident in Fig.~\ref{F:MAM:1}(b)
showing only vertical shift of individual master curves attributed to
an increasing stiffness when moving up individual scales. 

Finally, fitting the inverse of the homogenized master curve to a
particular chain of Maxwell units then allows for estimating the
macroscopic response of MAm to a given load limiting the material
symmetry to a macroscopic isotropy. The response of the homogenized
asphalt mixture to the applied remote shear strain labeled as ``Leonov
macro'' appears in Fig.~\ref{F:MAM:1}(a). Since derived from the
solution of a unit cell problem, a relatively good agreement with
these solutions has been expected. The difference is merely attributed
to the specific approximation of the homogenized master curve via
Dirichlet series.  It is fair to mention a considerable sensitivity of
the predictions to a particular choice of pairs of parameters (shear
modulus and retardation time) of the Maxwell units. Although
promising, to fully accept the proposed uncoupled multiscale
computational strategy will require further testing, both numerical
and laboratory. This topic is still widely opened. 

%\section{Macroscopic response - is coupled multiscale analysis needed?}\label{sec:MSA}
%%%%%%%%%%%%%%%%%%%%%%%%%%%%%%%%%%%%%%%%%%%%%%%%%%%%%%%%%%%%%%%%%%%%%%%%%%%%%%%%%%%%%%%%%%%%%%%%%%%%%%%%%%%%%%%%%%%%%%%%%%%%%

\section{Conclusions}\label{sec:con}
%%%%%%%%%%%%%%%%%%%%%%%%%%%%%%%%%%%%%%%%%%%%%%%%%%%%%%%%%%%%%%%%%%%%%%%%%%%%%%%%%%%%%%%%%%%%%%%%%%%%%%%%%%%%%%%%%%%%%%%%%%%%%
Although research interests on flexible pavements have been quite
intense in the past two decades, the field is still very much in
development and will certainly witness considerable activity in the
coming decade particular in connection to hierarchical modeling and
micromechanics. Within this framework, the present contribution
provides theoretical tools for the formulation of macroscopic
constitutive law reflecting the confluence of threads coming from
experimental work, image analysis, statistical mechanics and
traditional disciplines of micromechanics and the first order
computational homogenization. Here, the totally uncoupled multiscale
modeling approach is favored to enable an inexpensive analysis of real
world large scale structures. Since much of the considered is
primarily computational it would be audacious to brought this approach
directly to points of application without additional experimental
validation. Large scale experiments of rut depth measurements due to
moving wheel load are currently under way. The effect of
microstructure anisotropy, influence of the first stress invariant or
three-dimensional character of asphalt mixtures are just a few issues
which need to be addressed, yielding flexible pavements a fertile
field of future research.

\ack
%%%%%%%%%%%%%%%%%%%%%%%%%%%%%%%%%%%%%%%%%%%%%%%%%%%%%%%%%%%%%%%%%%%%%%%%%%%%%%%%%%%%%%%%%%%%%%%%%
{This outcome has been achieved with the financial support of the
Ministry of Education, Youth and Sports, project No. 1M0579, within
activities of the CIDEAS research centre.}
%%%%%%%%%%%%%%%%%%%%%%%%%%%%%%%%%%%%%%%%%%%%%%%%%%%%%%%%%%%%%%%%%%%%%%%%%%%%%%%%%%%%%%%%%%%%%%%%%

\appendix{
\section{The Mori-Tanaka predictions based on correspondence principle}
%%%%%%%%%%%%%%%%%%%%%%%%%%%%%%%%%%%%%%%%%%%%%%%%%%%%%%%%%%%%%%%%%%%%%%%%%%%%%%%%%%%%%%%%%%%%%%%%%
The correspondence principle as a tool allowing for interconversion of
linear viscoelastic response functions has already been mention in
Section~\ref{bsec:leonov}. Although not used in the present
contribution, this principle is reviewed here to fill in the list of
possible routes for deriving the homogenized viscoelastic response of
composites using simple averaging schemes. As an example, the
Mori-Tanaka method is adopted again to construct the homogenized
master curve of a two-phase composite medium composed of aligned
elastic circular fibers bonded to a viscoelastic mastic phase. This
approach thus allows direct comparison with the finite element
predictions obtained on the mortar scale.

Consider a homogenized relaxation loading path written as
\begin{eqnarray}
\Sigma_{xy}(t)&=&G_{\sf{hom}}(t)E_{xy},\label{eq:visco}\\
G_{\sf{hom}}(t)&=&\sum_{\mu=1}^{M}G_\mu^{\sf{hom}}\text{exp}\left(-\frac{t}{\theta_{\mu}^{\sf{hom}}a_{\sigma}^{\sf{hom}}a_T^{\sf{hom}}}\right),\label{eq:laplace:G}
\end{eqnarray}
where $M$ is the number of Maxwell units representing the homogenized
response, $a_{\sigma}^{\sf{hom}}=1$ in the viscoelastic regime,
$a_T^{\sf{hom}}=1$ for the reference temperature $T=40^0$C
(assumptions complying with the corresponding master curve shown in
Fig.~\ref{F:PHA:1}(b)) and $\theta_{\mu}^{\sf{hom}}$ are appropriate
relaxation times. Then, according to the correspondence principle the
Laplace-Carson transform of Eq.~\eqref{eq:visco} gives
\begin{eqnarray}
\Sigma_{xy}^*(p)&=&G_{\sf{hom}}^*(p)E_{xy}^*(p),\label{eq:laplace}\\
G_{\sf{hom}}^{*}(p)&=&\sum_{\mu=1}^{M}\frac{G_\mu^{\sf{hom}}}{(\displaystyle{\frac{1}{\theta_{\mu}^{\sf{hom}}}}+p)}.
\end{eqnarray}
Following~\cite{Lackner:ACPSEM:2007} provides Eq.~\eqref{eq:m-hom-MT}
in the Laplace space in the form
\begin{equation}
m^*_{\sf{MT}}=\frac{m_s^*m^*_m(k_m^*+2m^*_m)+k^*_mm^*_m(c_sm^*_s+c_mm^*_m)}{k^*_mm^*_m+(k^*_m+2m*_m)(c_sm^*_m+c_mm^*_s)},\label{eq:MTLAP}
\end{equation}
Since the volumetric response is assumed to remain elastic, recall Eq.~\eqref{eq:GLeonov1}, we get
\begin{eqnarray}
k_s^{*}(p)&=&\frac{m_s^{*}}{(1-2\nu_s)},\hspace{1.0cm}m_s^{*}\,=\,\frac{m_s}{p},\\
k_m^{*}(p)&=&\frac{m_m^{el}}{p(1-2\nu_m)},\hspace{0.55cm}m_m^{el}\,=\,\sum_{\mu=1}^N{G_{\mu}^m},\\
m_m^*(p)&=&\sum_{\mu=1}^{N}\frac{G_\mu^{m}}{(\displaystyle{\frac{1}{\theta_{\mu}^{m}}}+p)}.
\end{eqnarray}
%\begin{eqnarray}
%k_s^{*}(p)&=&\frac{m_s^{*}}{(1-2\nu_s)},\hspace{1.0cm}m_s^{*}\,=\,\frac{m_s}{p},\\
%k_m^{*}(p)&=&\frac{m_m^{el}}{p(1-2\nu_m)},\hspace{0.55cm}m_m^{el}\,=\,\frac{1}{J_{\mu=0}^m},\\
%m_m^*(p)&=&\frac{1}{p^2J_m^*(p)},\hspace{1.15cm}
%J_{m}^{*}(p)=\sum_{\mu=1}^{N}\frac{J_\mu^{m}}{p(p\tau_{\mu}^ma_{\sigma}^ma_T^m+1)},
%\end{eqnarray}
%
where $N$ is the number of Maxwell units approximating the behavior of
the mastic phase and $a_{\sigma}^m=a_T^m=1$ is again adopted to
guarantee compatibility with the results presented in
Section~\ref{bsec:leonov-mastic}. Once the Mori-Tanaka estimate
$G_{\sf{MT}}$ of the homogenized relaxation function
$G_{\sf{hom}}$ in the Laplace space are known, the desired coefficients
$G_{\mu}^{\sf{hom}}$, which enter Eq.~\eqref{eq:laplace:G}, are found by solving
the following minimization problem
\begin{equation}
E_G = \sum_{p_k}^{K}\left[G_{\sf{MT}}^*(p_k)-G_{\sf{hom}}^*(p_k)\right]^2\,=\,\text{min},\label{eq:min}
\end{equation}
thereby avoiding a cumbersome and rather unstable numerical
transformation of Eq.~\eqref{eq:MTLAP} back to time dependent
solution~\eqref{eq:laplace:G}. Solution strategies based on soft
computing~\cite{Ibrahimbegovic:2004,Kucerova:CAMES:2007,Kucerova:2007:PHD}
are typically employed to solve Eq.~\eqref{eq:min}.

Given the relationship between relaxation and compliance functions
\begin{equation}
pJ^*(p) = \frac{1}{pG^*(p)},\label{eq:GJ-LC}
\end{equation}
allows us to write a dual representation of Eq.~\eqref{eq:MTLAP} in
the form
\begin{equation}
p^2J_{\sf{MT}}^{*}(p)=\frac{1}{m_{\sf{MT}}^*}\,=\,
\frac{\displaystyle{\frac{k_m^*}{p^2J_m^*}}+\left(k_m^*+\displaystyle{\frac{2}{p^2J_m^*}}\right)\left(\displaystyle{\frac{c_s}{p^2J_m^*}}+c_mm_s^*\right)}
{\displaystyle{\frac{m_s^*}{p^2J_m^*}}\left(k_m^*+\displaystyle{\frac{2}{p^2J_m^*}}\right)+
\displaystyle{\frac{k_m^*}{p^2J_m^*}}\left(c_sm_s^*+\displaystyle{\frac{c_m}{p^2J_m^*}}\right)},\label{eq:MTLAP-J}
\end{equation}
where 
\begin{equation}
J_{m}^{*}(p)=\sum_{\mu=1}^{N}\frac{J_\mu^{m}}{p(p\tau_{\mu}^m+1)}.
\end{equation}
The homogenized coefficients $J{\mu}^{\sf{hom}}$ then follow from the
minimization of a dual error function
\begin{equation}
E_J = \sum_{p_k}^{K}\left[J_{\sf{MT}}^*(p_k)-J_{\sf{hom}}^*(p_k)\right]^2\,=\,\text{min}.\label{eq:min-J}
\end{equation}
or directly from Eq.~\eqref{eq:GJ-LC} providing the set of
coefficients $G{\mu}^{\sf{hom}}$ has been already found from the
primary formulation.

The resulting MT prediction of $J_{\sf{MT}}$ for the reference
temperature $T=40^0$C is compared in Fig.~\ref{F:LT}(b) with the PUC
solution on the mortar scale, recall Fig.~\ref{F:PHA:1}(b). Inability
of the MT method to capture the stiffening effect already mentioned in
Section~\ref{bsec:leonov-mortar} is evident in both the Laplace and
time space. It is interesting to point out that such a discrepancy
appears already for the elastic predictions if setting $m_m\rightarrow
0$ and $\nu_m\rightarrow 0.5$, in which case the binder behaves as an
incompressible fluid.

%%%%%%%%%%%%%%%%%%%%%%%%%%%%%%%%%%%%%%%%%%%%%%%%%%%%%%%%%%%%%%%%%%%%%%%%%%%%%%%%%%%%%%%%%%%%%%
\begin{figure}[ht]
\begin{center}
\begin{tabular}{c@{\hspace{1mm}}c}
\includegraphics[angle=0,width=6.75cm]{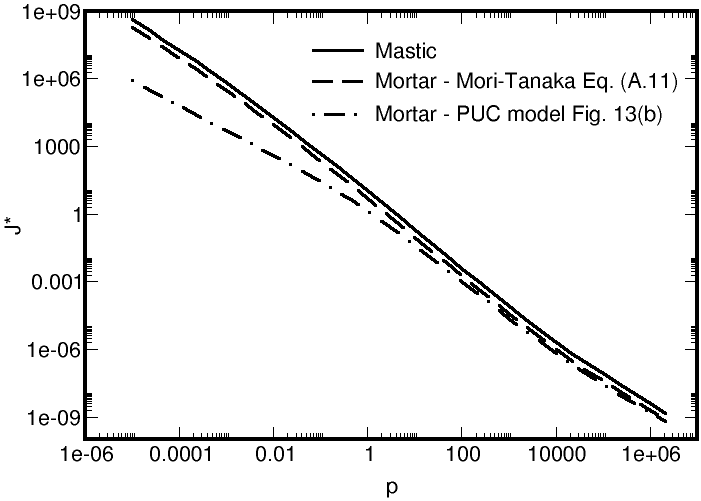}&
\includegraphics[angle=0,width=6.75cm]{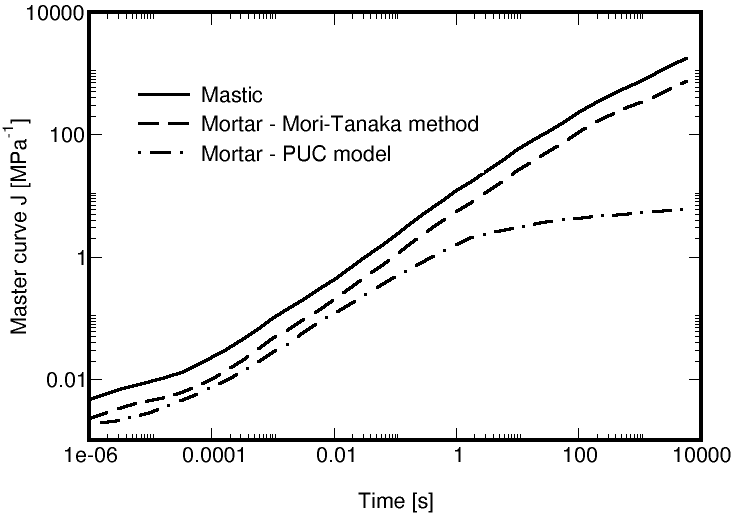}\\
(a)&(b)
\end{tabular}
\end{center}
\caption{(a) Laplace transform of the compliance function, (b) Master curve for reference temperature $T=40^0$C}
\label{F:LT}
\end{figure}
%%%%%%%%%%%%%%%%%%%%%%%%%%%%%%%%%%%%%%%%%%%%%%%%%%%%%%%%%%%%%%%%%%%%%%%%%%%%%%%%%%%%%%%%%%%%%%
}

\end{document}